\documentclass[]{aastex631}
\usepackage{natbib}

\defcitealias{A20}{A20}
\defcitealias{M20}{M20}
\defcitealias{M22}{M22}

\usepackage{multirow}
\usepackage{soul}
\shorttitle{ALMA HF-LBC-2021}
\shortauthors{Asaki et al.}
\graphicspath{{./}{figures/}}

\begin{document}

\title{
ALMA High-frequency Long Baseline Campaign in 2021: \\
Highest Angular Resolution Submillimeter Wave Images for 
the Carbon-rich Star R~Lep}

\author[0000-0002-0976-4010]{Yoshiharu Asaki}
\affiliation{Joint ALMA Observatory \\
        Alonso de C\'{o}rdova 3107, Vitacura 763 0355, 
        Santiago, Chile}
\affiliation{National Astronomical Observatory of Japan, \\
        Los Abedules 3085 Oficina 701, Vitacura 763 0414, 
        Santiago, Chile}
\affiliation{Department of Astronomical Science, School of Physical Sciences, \\
        The Graduate University for Advanced Studies, SOKENDAI,
        2-21-1 Osawa, Mitaka, Tokyo 181-8588, Japan}

\author[0000-0002-7675-3565]{Luke T. Maud}
\affiliation{ESO Headquarters,
        Karl-Schwarzchild-Str 2 D-85748 Garching, Germany}
\affiliation{Allegro, Leiden Observatory, Leiden University,
        PO Box 9513, 2300 RA Leiden, The Netherlands}

\author[0000-0001-8123-0032]{Harold Francke}
\affiliation{Joint ALMA Observatory \\
        Alonso de C\'{o}rdova 3107, Vitacura 763 0355, 
        Santiago, Chile}

\author[0000-0003-0292-3645]{Hiroshi Nagai}
\affiliation{National Astronomical Observatory of Japan, \\
        2-21-1 Osawa, Mitaka, Tokyo 181-8588, Japan}

\author[0000-0002-8704-7690]{Dirk Petry}
\affiliation{ESO Headquarters,
        Karl-Schwarzchild-Str. 2, 85748 Garching, Germany}

\author{Edward B. Fomalont}
\affiliation{National Radio Astronomy Observatory,
        520 Edgemont Road, Charlottesville, VA 22903, USA}

\author[0000-0001-9549-6421]{Elizabeth Humphreys}
\affiliation{Joint ALMA Observatory \\
        Alonso de C\'{o}rdova 3107, Vitacura 763 0355, 
        Santiago, Chile}
\affiliation{ESO Vitacura, Alonso de Cordova 3107, Santiago, Chile}

\author[0000-0002-3880-2450]{Anita M.S. Richards}
\affiliation{Jodrell Bank Centre for Astrophysics, Department of Physics and Astronomy, \\
                University of Manchester, M13 9PL, UK}

\author[0000-0002-4579-6546]{Ka Tat Wong}
\affiliation{
IRAM ARC node, Institut de Radioastronomie Millim\'{e}trique, 300 rue de la Piscine, F-38406 Saint-Martin-d'H{\`e}res, France}
\affiliation{
Theoretical Astrophysics, Department of Physics and Astronomy, Uppsala University, Box 516, SE-751 20 Uppsala, Sweden}

\author[0000-0002-2490-1079]{William Dent}
\affiliation{Joint ALMA Observatory \\
        Alonso de C\'{o}rdova 3107, Vitacura 763 0355, 
        Santiago, Chile}

\author[0000-0002-0465-5421]{Akihiko Hirota}
\affiliation{Joint ALMA Observatory \\
        Alonso de C\'{o}rdova 3107, Vitacura 763 0355, 
        Santiago, Chile}
\affiliation{National Astronomical Observatory of Japan, \\
        Los Abedules 3085 Oficina 701, Vitacura 763 0414, 
        Santiago, Chile}

\author{Jose Miguel Fernandez}
\affiliation{Lowell Observatory \\
        1400 W  Mars Hill Rd, Flagstaff, 
        AZ 86001, USA} 

\author[0000-0002-7287-4343]{Satoko Takahashi}
\affiliation{National Astronomical Observatory of Japan, \\
        2-21-1 Osawa, Mitaka, Tokyo 181-8588, Japan}
        
\author[0000-0001-5073-2849]{Antonio S. Hales}
\affiliation{Joint ALMA Observatory \\
        Alonso de C\'{o}rdova 3107, Vitacura 763 0355, 
        Santiago, Chile}
\affiliation{National Radio Astronomy Observatory,
        520 Edgemont Rd. Charlottesville, VA 22903, USA}



\begin{abstract}
The Atacama Large Millimeter/submillimeter Array (ALMA) was used in 2021 to image the carbon-rich evolved star R~Lep in Bands~8--10 (397--908~GHz) 
with baselines up to 16~km. The goal was to validate the calibration, using band-to-band (B2B) phase referencing with a close phase calibrator J0504-1512, $1.^{\circ}2$ from R~Lep in this case, and the imaging procedures required to obtain the maximum angular resolution achievable with ALMA. Images of the continuum emission and the hydrogen cyanide (HCN) maser line at 890.8~GHz, from the $J$=10--9 transition between the ($11^{1}0$) and ($04^{0}0$) vibrationally excited states, achieved angular resolutions of 13, 6, and 5~mas in Bands~8--10, respectively. Self-calibration (self-cal) was used to produce ideal images as to compare with the B2B phase referencing technique. The continuum emission was resolved in Bands~9 and 10, leaving too little flux for self-cal of the longest baselines, so these comparisons are made at coarser resolution. Comparisons showed that B2B phase referencing provided phase corrections sufficient to recover 92\%, 83\%, and 77\% of the ideal image continuum flux densities. The HCN maser was sufficiently compact to obtain self-cal solutions in Band~10 for all baselines (up to 16~km). In Band~10, B2B phase referencing as compared to the ideal images recovered 61\% and 70\% of the flux density for the HCN maser and continuum, respectively.
\end{abstract}


\section{
  Introduction
}\label{sec:01}
\subsection{
  Atacama Large Millimeter/submillimeter Array evolution to high angular resolution
}\label{sec:01-01}

The Atacama Large Millimeter/submillimeter Array (ALMA), located on the Chajnantor plain of the Atacama desert in Chile, is a large and sensitive radio array designed to image celestial sources at millimeter to submillimeter wavelengths. A detailed description of the observatory at the time of the observations discussed here can be found in 
\cite{Cortes2022}.
The telescope system consists of two arrays and one single dish cluster. One of the arrays, designated as the 12~m array, contains up to fifty 12~m diameter antennas, and
provides a range of angular resolutions by reconfiguring 
the antenna locations within 10 different configurations. 
The approximate angular resolution of a radio array is roughly 
$\lambda / B_{\mathrm{max}}$~rad, where $\lambda$ is the observing 
wavelength, and $B_{\mathrm{max}}$ is the longest baseline length projected 
to the target source in the array\footnote{In the case of a large array 
with an antenna distribution like ALMA, a more accurate estimate of angular resolution replaces
$B_{\mathrm{max}}$ with $L_{80}$, i.e. the baseline length corresponding 
the 80th percentile of the baseline length distribution, see 
\cite{Cortes2022}.}. 
The highest angular resolution possible with the 12~m array is achieved in Configuration~10, with the longest baseline length of 16.2~km.  

Relatively compact configurations were available in early ALMA observing cycles, while progressively longer baselines or higher frequencies have been made available during the period 2013--2019 based on capabilities confirmation.
In 2013, the first relatively long baselines of 2.7~km were coordinated by putting three remote antennas located far from the regular 12~m array to make science verification 
experiments at 350~GHz (Band~7) and 650~GHz (Band~9) 
\citep{2014A&A...572L...9R}. 
In 2014, the 12~m array was used for the first time in the most extended configuration with the longest baselines of 16~km using 
Bands~3 (84--116~GHz), 
4 (125--163~GHz), 
6 (211--275~GHz), and 
7 (275--373~GHz) 
\citep{2015ApJ...808L...1A,
2015ApJ...808L...2A,
2015ApJ...808L...3A, 
2015ApJ...808L...4A, 
2015A&A...577L...4V,
2015ApJ...808...36M,
2016A&A...590A.127W}. 

On the other hand, ALMA's longest baseline capability had not been opened to users of the highest frequency bands (Bands~8--10). Using a phase reference calibrator as close as possible to the target was found to be critical in relatively long baseline ($>$3.6~km), 
high-frequency (HF) observations from the interferometric phase stability point of view. 
An approximate guideline is that the phase calibrator should be within $4^{\circ}$ at Band~7 for the most extended configurations and less distant at Bands~8 (385--500~GHz), 9 (602--720~GHz) and 10 (787--950~GHz) 
\citep[][hereafter M20]{M20}. 
However, the sky number density of suitable phase calibrators decreases at higher frequencies  since the majority of phase calibrators are quasars, which become fainter at higher frequencies while the 
atmospheric attenuation for the submillimeter waves becomes larger
\citep[][hereafter A20]{A20}. 

In order to overcome the difficulties in finding a close phase calibrator in the HF bands, ALMA needed to wait for the validation of the novel phase calibration technique using band-to-band (B2B) phase referencing 
\citep{2016SPIE.9906E..5UA}. 
With this calibration method, a phase calibrator is observed at a lower frequency, and then the phase 
corrections are derived by scaling the solutions up to the target frequency 
\citep[e.g.,][]{1998RaSc...33.1297A,
2009arXiv0910.1159D,
2010ApJ...724..493P}. 
In using B2B phase referencing, the number of usable phase calibrators is increased and it becomes possible to find a calibrator closer to the science target 
\citep[typically $<3^{\circ}$,][]{2023ApJS..267...24M}
compared to the case where the calibrator is required to be at the same frequency as the science target.

The initial High-frequency Long Baseline Campaign 
in 2017 (HF-LBC-2017) was arranged to explore 
B2B phase referencing 
using the 12~m array with up to 16~km baselines  
in Bands 7--9 
\citep{A20,
2020AJ....160...59A,
M20,
M22}.
In parallel, verification of an antenna fast switching to alternate scans 
between the phase calibrator and target as quickly as possible 
during an observation was being tested as to mitigate the fast visibility phase variations due to atmospheric phase fluctuations.
The switching cycle time ($t_{\mathrm{swt}}$) 
is defined as a time interval between two successive phase calibrator scans: such that phase referencing corrects 
the timescale phase 
variations between the phase calibrator and the target scans longer than 
that with the timescale of $t_{\mathrm{swt}}/2$
\citep{1999RaSc...34..817C}. 
Thus, fast switching acts to minimize achievable and correct phases over short timescales. 
The experiments and tests of the ALMA fast switching in 
\cite{M22} (hereafter M22) 
suggest that $t_{\mathrm{swt}}$ should be no longer than 60~s 
for ALMA long baseline observations in Bands~8--10, 
with the proviso that the phase calibrator is also close to
the target. 

In the 2019 High-frequency Long Baseline Campaign 
(HF-LBC-2019), a combination of B2B phase referencing and fast switching was extensively tested in Bands~8, 9, and 10, and the impact of the phase calibrator separation angle was investigated. 
In the tests, bright quasars were used as both the phase calibrator and target sources, and images with an angular resolution of the minor axis of the synthesized beam of $\sim 5$~mas were achieved 
\citep{2023ApJS..267...24M}. 
HF-LBC-2019 demonstrated that either  
standard phase calibration (in-band phase referencing) or B2B phase referencing 
produced {\it good image quality} 
(Section~\ref{sec:02-02}) 
when using a sufficiently close phase calibrator. 
For example, for Band~10 and 16~km baselines, 
the visibility phase residual rms of $\leq$1~rad 
could be achieved by selecting a phase calibrator with a separation angle of ${\leq}1^\circ$ and provided that the atmospheric phase stability was $\leq$30$^\circ$ as measured over $t_{\mathrm{swt}}$ 
\citep{2023ApJS..267...24M}.

Following the positive results of HF-LBC-2019, the High-Frequency Long Baseline Campaign in 2021 (HF-LBC-2021) was organized to observe a compact target with an extended array configuration, including 16\,km baselines using the same observing and data processing procedures as in users' observations.
We selected R~Lep, a carbon-rich, mass-losing asymptotic giant branch (AGB) star, of which the continuum emission was thought to be compact from low angular resolution ALMA observations in Bands~6 and 7 \citep{2020A&A...640A.133R}, and Band~10 \citep{Wong2019}.
It has a sufficiently close phase calibrator (1.$^\circ$2) when using B2B phase referencing, and the star has relatively strong continuum emission and a bright, narrow maser line in Band~10. R~Lep was observed in three experiments, one in each of Bands~8--10. 

The HF-LBC-2021 experiments with R~Lep as the science target are described as follows:
Section~\ref{sec:02} 
provides an overview and the technical background of the experiments. 
Section~\ref{sec:03}  
briefly details the data reduction and calibration procedure.
Section~\ref{sec:04} 
presents the observation results for the target source, while a discussion of these results follows in 
Section~\ref{sec:05}. 
Phase calibrators are likely to show resolved structure at high angular resolution, and 
Section~\ref{sec:05} 
discusses the implications if a calibrator turns out to be resolved. Finally, we summarize the paper in  
Section~\ref{sec:06}. 

\subsection{
  The Carbon-rich Star R~Lep
}\label{sec:01-02}

R~Lep is a Mira-type variable with a period 
of 445 days \citep{2006SASS...25...47W} at a distance of $471_{-64}^{+88}$~pc, based on the estimate by 
\citet{2022A&A...667A..74A} 
using the third \emph{Gaia} data release 
\citep[DR3;][]{2023A&A...674A...1G}. 
R~Lep has an apparently spherical shape and a diameter of $15.20 \pm 0.20$~mas at a pulsation phase of 0.20, in the near-infrared $K$-band (2.0 $-$ 2.4~$\mu$m), as measured by the Very Large Telescope Interferometer 
\citep[VLTI; ][]{2005ESASP.560..651H}. 
We adopt a systemic velocity of 11.5~km~s$^{-1}$ and an outflow velocity of 19~km~s$^{-1}$ based on the CO observations of 
\citet{2020A&A...640A.133R}. 
The coordinates of the star are ($04^{\mathrm{h}}59^{\mathrm{m}}36^{\mathrm{s}}.35$, ~$-14^{\circ}48'22''.5$) in the International Celestial Reference System (ICRS) at J2000. 
The continuum emission of AGB stars at radio and (sub)millimeter wavelengths is consistent with the {\it radio photosphere} model, in which the dominant opacity is caused by free--free interactions between electrons and neutral H and H$_2$, and the spectral index is roughly 2 in the radio domain 
\citep{1997ApJ...476..327R,
2016A&A...586A..69P}.
The continuum flux densities of R~Lep in Bands~6 (224~GHz) and 7 (338~GHz), as measured from the images provided by
\citet{2020A&A...640A.133R}
at the CDS\footnote{Strasbourg Astronomical Data Center, \url{https://cdsarc.u-strasbg.fr/viz-bin/cat/J/A+A/640/A133}}, are 16 and 27~mJy, respectively, implying a spectral index of ${\sim}1.3$. The flux density at Band~10 (890~GHz) is expected to be ${\gtrsim}100$~mJy.

Mass loss from AGB stars builds up a circumstellar envelope (CSE) rich in molecules and dust. In using different molecular species we can probe the CSE, and therefore the mass-loss process, at different radii from the central stars.
Hydrogen cyanide (HCN) is one of the most abundant molecules in carbon-rich CSEs and is known 
to exhibit maser action in many transitions in the (sub)millimeter domain 
\citep[e.g.,][]{
1987A&A...176L..24G,
Bieging2001,
2014MNRAS.440..172S,
Menten2018},
especially in the vibrationally excited states of the bending mode
\citep[see Table A.1 of][for a list of known HCN maser from C-rich CSEs to date]{2022A&A...666A..69J}.
There are two submillimeter HCN masers at 804.75 and 890.76 GHz, first detected in C-rich CSEs by
\citet{2000ApJ...528L..37S}
and
\citet{2003ApJ...583..446S},
respectively, using the Caltech Submillimeter Observatory (CSO) 10.4~m telescope on Maunakea, which are observable in Band~10 with ALMA. The HCN maser targeted in R Lep is that at 890.7607~GHz, from the $J=10-9$ transition between the ($11^{1}0$) and ($04^{0}0$) vibrationally excited states with an upper-level energy above 4260~K 
\citep{1967PhLA...25..489H,2014MNRAS.437.1828B}.

Circumstellar astronomical masers have only been resolved on milliarcsecond scales previously around O-rich stars. SiO masers occur within a few stellar radii, where we expect to find the HCN masers. SiO masers emanate from clumps with a typical extent of $\sim$1~au \citep{2018ApJ...869...80A} while the individual, beamed spots are much smaller (as established by component fitting). At the distance of R~Lep, 1~au corresponds to an angular size of 2.1~mas, which cannot be resolved directly even with ALMA's highest angular resolution of 5~mas. A whole SiO maser shell would have a diameter $\lesssim$50 mas at the distance of R~Lep 
\citep[e.g.,][]{2004A&A...414..275C}, 
just within the ALMA Band 10 maximum recoverable scale. 

\cite{Wong2019} 
reported the first ALMA imaging of bright HCN masers toward C-rich stars including R~Lep at an angular resolution of $\sim 0.''1$, and found that the spatial structure of the HCN masers is generally not well resolved, confirming previous predictions that these masers should arise very close to the star 
\citep{2003ApJ...583..446S}.
This gives us some a~priori knowledge about the probable properties of target images in these long baseline experiments. The bright and relatively compact maser emission also makes it easy to perform self-calibration (self-cal), which is helpful in assessing the image coherence (see Section~\ref{sec:03-05-02}).
Therefore, R~Lep is an excellent target for use in fully validating ALMA's long baseline capability in Band~10.

\section{
  Experiments
}\label{sec:02}
\subsection{
  Overview
}\label{sec:02-01}

The ALMA long baseline imaging capability in Bands 8--10 was demonstrated in HF-LBC-2021 by observing R~Lep. In the experiments, the quasar J0504-1512, whose separation angle from R~Lep is $1.^{\circ}2$, was selected as the phase calibrator. One experiment run was performed in each of Bands~8--10 and the parameters for the three experiments are listed in  
Table~\ref{tbl:01}.  
Note that the experiment codes B-08--B-10 indicate the observing frequencies in Bands~8--10, respectively. 
Table~\ref{tbl:02} 
lists the other calibrators and check sources (quasars used to independently assess the data quality). A 10~minute bandpass scan on a bright quasar was placed at the beginning of the experiment as shown in 
Figure~\ref{fig:01}. 
This scan determines the amplitude and phase change as a function of frequency for each of the HF spectral windows, and is then applied to all scans in the experiment, as per standard bandpass calibration for all ALMA data. 

In Band~8 we set up two HF spectral windows each with a bandwidth of 1.875~GHz (frequency width per channel of 976.56~kHz) and the other two HF spectral windows with a bandwidth of 2~GHz (frequency 
width per channel of 15.625~MHz), as listed in 
Table~\ref{tbl:03}. 
In Bands~9 and 10, we used eight HF spectral windows with a bandwidth of 1.875~GHz (frequency width per channel of 976.56~kHz), made possible with $90^{\circ}$ Walsh phase switching to enable sideband separation in the double sideband receiving systems 
\citep{ALMA_MEMO_537}. 
Each spectral window was observed in the two linear polarization pairs, called $XX$ and $YY$. In Band~10, one of the spectral windows covered the HCN maser at 890.7607~GHz in the $J$=10--9 transition between the ($11^{1}0$) and ($04^{0}0$) vibrationally excited states.  

Table~\ref{tbl:03}
also lists low-frequency (LF) spectral windows used for B2B phase referencing. They were all acquired with single sideband receiving systems. In B-09 and B-10, two of the four LF spectral windows to observe were not available due to instrumental problems. These spectral windows were neither included in the data reduction, nor listed in 
Table~\ref{tbl:03}. 

\subsection{
  Goals of the experiments
}\label{sec:02-02}

The goals of our R~Lep observations are to confirm that the observation setup, and calibration and image procedures for ALMA HF long baseline observations can achieve accurate and representative images of the science target source while following the below scenario:
\begin{list}{}
{
  \setlength{\itemindent}{0mm}
  \setlength{\parsep}{0mm}
  \setlength{\topsep}{3mm}
}
\item[(1)]
Using B2B phase referencing,
\item[(2)]
Using a close phase calibrator,  
\item[(3)]
Fast switching between the target and a phase calibrator, and
\item[(4)]
Assessing phase stability prior to execution using a ``Go/NoGo'' check.
\end{list}{}

In the case that the visibility phase includes random phase noise, the coherence factor of the visibilities, $\eta$, can be estimated as follows: 
\begin{eqnarray}
\label{eq:01}
\eta &=& 
\exp{[-\sigma_{\mathrm{\Phi}}^2 / 2}],
       \end{eqnarray}
where $\sigma_{\mathrm{\Phi}}$ is the standard deviation of the interferometric phase noise
\citep[][Section~13.1.6]{2017isra.book.....T}. 
In general, synthesized images generated by Fourier transforming point-source visibilities suffer from an image peak degradation equivalent to Equation~(\ref{eq:01}) due to random phase noise that we characterize as a ratio of the obtained image peak to that of the true brightness distribution convolved with the synthesized beam. We refer to this as the image coherence. In our previous studies image structure defects are minimal to negligible for point sources with a 70\% image coherence and under these conditions, we therefore expect images of science targets to be accurate and representative 
(\citetalias{M20}, \citetalias{M22}).

Although R~Lep may have an extended structure, we also apply these criteria to the HF-LBC-2021 experiments, in that we ensure good phase stability before observing and maintain a low residual phase rms for the target after B2B phase referencing by using a close phase calibrator. 
For R~Lep we assume that the beam-convolved true brightness distribution can be retrieved by self-cal as discussed in 
Section~\ref{sec:03-05}, 
and thus we can measure the image coherence by comparing an image made after B2B phase referencing with those made after self-cal.
In the following subsections, the application of the above techniques and criteria in order to achieve an image coherence of $\geq 70$\% are described. 

\subsubsection{
  B2B phase referencing
}\label{sec:02-02-01}

For most HF experiments, especially in Bands~9 and 10, pointlike quasars lying within a suitable angular separation from the target source will not be always bright enough to provide robust phase solutions. However, since the flux densities of quasars often increase at lower frequencies, it is possible to apply the solutions derived at lower frequencies to the higher frequency data using the B2B phase referencing. As detailed in 
\citet{2023ApJS..267...24M}, 
irrespective of using in-band or B2B phase referencing, the calibrator separation angle to the target should be $\sim$1$^\circ$ in Bands~9 and 10 before using the longest baseline configuration. Only when using B2B phase referencing for R~Lep does the quasar J0504-1512, at $1^{\circ}.2$, meet the separation angle requirement. In the following descriptions, the observing band used for the target is referred to as an HF band with the HF representative frequency of $\nu_{\mathrm{_{\mathrm{HF}}}}$. The lower frequency band used to observe the phase calibrator is referred to as an LF band with the LF representative frequency of $\nu_{\mathrm{_{\mathrm{LF}}}}$. The LF band spectral setup is automatically determined by the ALMA observatory control software (ALMA online software) to minimize the system overheads incurred when frequency switching between the HF and LF bands to $\sim 2$~s by using a harmonic frequency switching method
(\citetalias{A20}; \citealt{2012SPIE.8452E..16S}). 

In the data reduction, the antenna-based phase of the phase calibrator at $\nu_{\mathrm{_{\mathrm{LF}}}}$ with respect to a reference antenna is scaled to $\nu_{\mathrm{_{\mathrm{HF}}}}$ by multiplying by the frequency ratio, $R = \nu_{\mathrm{_{\mathrm{HF}}}} / \nu_{\mathrm{_{\mathrm{LF}}}}$. This scaling is relevant if the variable phases are associated with delays in each antenna path, such as those caused by the troposphere, antenna position error and other common instrumental paths between the antenna and the correlator.
See 
\citetalias{A20} 
for more details. 

There is also an additional instrumental phase offset, independent of the frequency scaling, that exists between the HF and LF spectral windows. 
These phase offsets (one for each spectral window) are calibrated by observing a strong quasar, called the Differential Gain Calibration (DGC) source, while alternating very rapidly between the HF and LF bands such that atmospheric variations are corrected and the remaining phase offset can be determined
\citepalias{A20}.
For each of the three R~Lep experiments, four DGC sequences were used to determine if the spectral window phase offsets were constant over each experiment. The observing sequence of B-09 is depicted in 
Figure~\ref{fig:01}. 
As noted by 
\citetalias{A20}, 
the DGC source does not need to be close to the science target, although it must be bright so that the frequency switching scans can be made faster than the switching cycle between the target and phase calibrator. In the experiments, the DGC source used was J0522-3627, which has an angular separation from the target of $22^{\circ}.3$, and a flux density of $>4$~Jy in the HF bands 
(see Section~\ref{sec:03-02}), 
and so it is sufficiently strong. However, it is not a perfect point source 
\citep{2016A&A...586A..70L}, 
and the possible effect of its structure on obtaining the spectral window phase offsets is discussed in 
Section~\ref{sec:05-02}. 

\subsubsection{
  Antenna fast switching and the Go/NoGo check
}\label{sec:02-02-02}

Short timescale atmospheric 
phase fluctuations affect the phase stability even in 
low precipitable water vapor (PWV) conditions
\citep{
2016SPIE.9906E..5UA,
2017PASP..129c5004M,
2017A&A...605A.121M,
2023arXiv230408318M}.
In order to reduce the fast phase variations in ALMA phase referencing observations, while retaining 
adequate target on-source time, 
we chose the switching cycle time $t_{\mathrm{swt}}$ to be 50~s. 
The target scan length was $\sim 31$~s and the phase calibrator 
scan length was $\sim 13$~s, with an overhead of $\sim 6$~s 
to change the antenna pointing and receiver band. 
We imposed the nominal ALMA limits  on the amount of PWV of 0.91, 0.66, and 0.47~mm for the experiments in Bands~8--10, 
respectively.  

Just before each experiment, we conducted a 120~s short observation of a bright quasar, J0423-0120, to check the atmospheric phase stability by measuring the phase rms at an arbitrary frequency $\nu_{\mathrm{_\mathrm{GNG}}}$ available at that moment. This short observation is referred to as a Go/NoGo check in this paper. We derived the phase rms at $\nu_{\mathrm{_\mathrm{HF}}}$ by scaling by the ratio between $\nu_{\mathrm{_\mathrm{GNG}}}$ and $\nu_{\mathrm{_\mathrm{HF}}}$. 
Note that ALMA operators monitor visibility phases corrected with solutions derived from Water Vapor Radiometer (WVR) measurements 
\citep{2013A&A...552A.104N}
using the ALMA online calibration software {\sc telcal} 
\citep{2011ASPC..442..277B}.
In general, the phase rms is proportional to the square root of the observation interval
\citepalias{M22}. 
The B2B phase referencing experiments have a switching cycle time of $\sim$50~s, giving a time scaling factor of $[120\ {\mathrm{s}}/50\ {\mathrm{s}}]^{0.5}=1.54$. Using the phase rms of the Go/NoGo check, we can predict the image coherence from a modified coherence factor $\eta'$ based on 
Equation~(\ref{eq:01}) 
as follows: 
\begin{eqnarray}
\label{eq:02}
\eta' &=& 
\exp{[-(\nu_{\mathrm{_\mathrm{HF}}}
       /\nu_{\mathrm{_\mathrm{GNG}}})^2
       (\sigma_{\mathrm{_{\mathrm{GNG}}}}/1.54)^2 / 2}], 
\end{eqnarray}
where $\sigma_{\mathrm{_{\mathrm{GNG}}}}$ is the median phase rms using 
the longest quartile of baseline lengths
from the Go/NoGo check at $\nu_{\mathrm{_{\mathrm{GNG}}}}$.
Given our aim to achieve an image coherence of $\geq 70$\%, we required
($\nu_{\mathrm{_\mathrm{HF}}} / \nu_{\mathrm{_\mathrm{GNG}}}) \sigma_{\mathrm{_{\mathrm{GNG}}}}$ 
$\leq 0.8$~rad ($=45^{\circ}$) for the longest quartile of baseline lengths 
\citepalias{M22}. 

Figure~\ref{fig:02} 
shows the phase rms as a function of the baseline length of the Go/NoGo checks, and 
Table~\ref{tbl:04} 
lists the weather conditions obtained from the Go/NoGo checks. The modified coherence factors calculated from 
Equation~(\ref{eq:02}) 
are 95\%, 85\%, and 77\% for the conditions of B-08--B-10, respectively. Further discussions about the relation between the Go/NoGo check and the actual image coherence are presented in 
Section~\ref{sec:05-01-02}.

\section{
  Data reduction
}\label{sec:03}

The data reduction was carried out using custom data reduction 
scripts generated by the ALMA calibration script generator 
\citep{2014SPIE.9152E..0JP}\footnote{
The utility module of the script generator can be downloaded at the 
following link: https://doi.org/10.5281/zenodo.7502159}
based on 
the Common Astronomy Software Applications 
\citep[CASA;][]{2022PASP..134k4501C}. 
As per standard procedures, the ALMA cross-correlated data, or in other words, Archival Science Data Models were converted to a {\sc CASA} MeasurementSet (MS). Thereafter, we applied the conventionally derived corrections 
such as WVR phase corrections
\citep{2013A&A...552A.104N,
2017PASP..129c5004M}, 
any updated antenna positions, and HF bandpass calibration to the MSs. 
In all such experiments, it is important that accurate antenna positions are used 
\citep{2016SPIE.9914E..2LH, 2023ApJS..267...24M}.
For the LF bandpass calibration, we made use of the LF DGC scans of J0522-3627 as listed in 
Table~\ref{tbl:02}.

\subsection{
  Phase calibration
}\label{sec:03-01}

There are three stages in our data reduction for B2B phase referencing. The first stage is the elimination of the instrumental phase offsets between the LF spectral windows; the second is the estimation of the instrumental phase offset for the HF spectral windows with respect to the LF band; and the final stage is applying the LF phase calibrator solutions using B2B phase referencing to correct the HF target phases. 
At the first stage (corresponding to step (a) in \citetalias{A20}, Figure~3), we obtained time-averaged LF band phase solutions for the DGC source J0522-3627 for each LF spectral window for each polarization pair. This set of DGC solutions is referred to as the LF DGC offset. For B-08, there are four LF (Band~4) spectral windows, each with two polarization pairs, and therefore we obtained in total eight LF DGC offsets (four spectral windows ${\times}$ two polarization pairs). On the other hand, for B-09 and B-10, there are two LF spectral windows available (Bands~4 and 7, respectively); therefore, we obtained four LF DGC offsets for each band. In the next step, the LF DGC offsets were subtracted from the corresponding LF spectral windows (steps (b) and (e) in \citetalias{A20}, Figure~3).

In the second stage, we derived time-dependent phase solutions for the DGC at $\nu_{\mathrm{_{\mathrm{LF}}}}$ for each scan. In B-09 and B-10 the solutions were averaged for both LF spectral windows to improve the signal-to-noise ratio (S/N). 
These solutions are referred to as the LF DGC solutions. The solutions were interpolated in time and extrapolated in frequency (by multiplying by the frequency ratio $R$), and applied to the HF DGC spectral windows,  to calibrate short timescale atmospheric phase fluctuations in the HF DGC scans (step (c) in \citetalias{A20}, Figure~3). 
This calibration essentially removes the temporal phase fluctuations and leaves only the offset between LF and HF spectral windows. Thus, in applying these corrections, we derive a time-averaged solution including scans per group of the HF DGC repeats (step (d) in \citetalias{A20}, Figure~3), producing eight and 16 HF DGC offsets in Band~8, and Bands~9 and 10, respectively, one
for each spectral window for each polarization pair. 

At the final stage, we derived per scan phase solutions for the LF phase calibrator J0504-1512. For B-09 and B-10, we combined the two LF spectral windows to improve the S/N, while for B-08 the individual spectral windows were solved separately. The HF DGC offsets were applied to the corresponding HF spectral windows to remove the instrumental phase offset, and thereafter the application of the scaled-up scan-based LF phase calibrator solutions are used to calibrate the HF target phases. The LF phase solutions were interpolated in time and extrapolated in frequency by multiplying $R$ 
(step (f) in \citetalias{A20}, Figure~3).

\subsection{
  Amplitude calibration
}\label{sec:03-02}

The amplitude calibration procedure for the experiments followed the ALMA standard amplitude calibration flow 
\citep{Cortes2022}. 
As displayed in 
Figure~\ref{fig:01}, 
$T_{\mathrm{sys}}$ measurement scans in the HF band were inserted roughly every 8 minutes and the derived corrections were applied to the data. Although the LF $T_{\mathrm{sys}}$ measurement is needed neither for the HF target phase nor amplitude calibrations, LF band amplitude correction with $T_{\mathrm{sys}}$ was also made for checking the synthesized images of the LF calibrators. 

The amplitude time variation of the HF spectral windows was determined using the DGC sequences in the HF band. The flux density scaling of R~Lep was finally made by referring to the HF flux density of J0522-3627, whose Band~3 and 7 flux densities have been measured in ALMA's grid source monitor 
\citep{
2014Msngr.155...19F,
ALMA_MEMO_599,
2021ApJS..256...19F}. 
Assuming that the flux density is proportional to $\nu^{\alpha}$, where $\nu$ is the observing frequency and $\alpha$ is the spectral index determined between Bands~3 and 7, the flux density of J0522-3627 extrapolated to Bands~8--10 was 4.481, 4.390, and 4.137~Jy at 405, 667, and 896~GHz, respectively. The obtained spectral index, computed flux densities of J0522-3627, and the measurement epochs are listed in 
Table~\ref{tbl:05}. The ramifications of the slight structure in J0522-3627 are discussed in 
Section~\ref{sec:05-02}.

\subsection{
  Visibility amplitude analysis of the continuum source
}\label{sec:03-03}

The top panels of 
Figure~\ref{}
show the ($u$,~$v$) coverage of the R~Lep target scans normalized by the wavelength. The maximum ($u$,~$v$) distance is 20, 36, and 48~M$\lambda$ in Bands~8--10, respectively, corresponding to angular scales of 10, 5.8, and 4.3~mas.  
Note that the 80th percentile of the ($u$,~$v$) distance is 9.4, 19, and 25~M$\lambda$ in Bands~8--10, respectively, corresponding to angular scales of 22, 11, and 8.3~mas.  
These divergences between the maximum and 80th percentile ($u$,~$v$) distance are caused by non-uniformly distributed ($u$,~$v$) coverage.  
In the following synthesis imaging using {\sc casa} {\tt tclean} we set the customary weighting of visibility samples (``Briggs'' robustness parameter of 0.5) to optimize ALMA imaging fidelity.

The bottom panels of 
Figure~\ref{}
show the calibrated visibility amplitude of the continuum emission as a function of ($u$,~$v$) distance. 
The first NULL point in the visibility amplitude appears at $\sim 15$~M$\lambda$ in Bands~8 and 9, while it is not as clearly seen in Band~10 due to the higher noise level relative to the other bands. Beyond this point, the amplitude remains low in Bands~8 and 9, and probably also in Band~10. Assuming that the morphology of the brightness distribution of the continuum source of evolved stars is well represented by a uniform disk at these frequencies
\citep{2018AJ....156...15M, 
2019A&A...626A..81V},
we fitted the visibilities to a disk model using {\sc casa} {\tt uvmodelfit}. The fitting results are listed in 
Table~\ref{tbl:06}. 
The continuum emission has a more extended distribution of $14-18$~mas than the synthesized beam sizes. This will be discussed in 
Section~\ref{sec:03-05-01}
in relation to self-cal.

\subsection{
  Imaging for R~Lep in the HF Bands
}\label{sec:03-04}

After the phase and amplitude calibration including B2B phase referencing, images were synthesized for R~Lep with an imaging area of $512 \times 512$~mas$^{2}$ with a pixel size of 1~mas, centered at ($04^{\mathrm{h}}59^{\mathrm{m}}36^{\mathrm{s}}.3590$, ~$-14^{\circ}48'22''.531$) in ICRS. {\sc casa} {\tt tclean} was used for imaging with a Briggs robustness parameter of 0.5 and a small cleaning box around the compact source at the center of the image. 
For the continuum images in Bands~8 and 9, the entire frequency range was used, whereas for Band~10 the known frequency range of the HCN maser emission, between 890.664 and 890.860~GHz, was excluded. 

For the Band~10 HCN maser, we subtracted the continuum emission from the spectral channels containing the HCN maser line using a linear fit to the calibrated visibilities in the surrounding channels in the corresponding spectral window (BB-4 LSB). 
The frequency axis was adjusted to the local standard of rest Kinematic 
(LSRK) frame with respect to the rest frequency of the HCN maser line. Those two procedures were made at once using {\sc casa}  {\tt mstransform}. An image cube of the HCN maser emission was made covering  100 channels, centered at 890.742~GHz, each with a frequency width of 976.562~kHz. This corresponds to a velocity width of $\sim 0.3$~km~s$^{-1}$. 

\subsection{
  Self-cal
}\label{sec:03-05}

In order to estimate the percentage of image coherence achieved using B2B phase referencing, we need to compare these images with the best possible target images, achieved using self-cal. The introduction of self-cal is covered by 
\citet{1984ARA&A..22...97P} 
(see also 
\cite{2022PASP..134k4501C} 
and references therein).  Self-cal uses CLEAN components from a phase-referenced image as a model. Their Fourier transform is compared with the observed visibilities to derive corrections (mainly to the phase variations caused by the troposphere during the target source scans), that when applied can lead to an improved image which may in turn be used in further rounds of self-cal as the coherence is step-wise improved.  
\citet{2018arXiv180505266B} 
and 
\citet{2022arXiv220705591R} 
describe recent implementations of self-cal for ALMA data reduction.
The measured image 
coherence is described in detail in 
Section~\ref{sec:05}. 

\subsubsection{
  Continuum self-cal
}\label{sec:03-05-01}

The CLEANed continuum images of R~Lep obtained after applying B2B phase referencing were used to provide the CLEAN component models to the ``MODEL'' column of the MSs for one round of phase self-cal. 
The target scan length (25$-$30~s on average) was used as the solution interval. We assumed that the continuum emission is unpolarized and, within each ALMA band, there are negligible changes in its structure  and angular extent, such that we combined the $XX$ and $YY$ polarization pairs and all the spectral windows to increase the S/N. 
In all cases, when applying the self-cal corrections, if the solution for a particular antenna and scan had failed, the corresponding data were flagged.

Self-cal decomposes per-baseline solutions into corrections applied per antenna, so a minimum S/N (usually 3) 
is required for each antenna, per solution interval. In Band~8, the continuum emission extends over only a factor of $\sim 1.5$ of the synthesized beam, so there was good S/N on all baseline lengths, and accurate phase self-cal solutions were obtained for almost all antennas and scans. A negligible amount of data was flagged due to failed solutions so
the synthesized beam sizes are very similar before and after phase self-cal.  

For Bands~9 and 10 the continuum source extends over 3--5 synthesized beams and the S/N on baselines to the most distant antennas was too low to provide calibration solutions in self-cal: good solutions were only obtained for the antennas contributing shorter baselines. In applying the corrections, much of the data were flagged due to failed solutions, so this led to a lower angular resolution for the images made after self-cal as the longest ($u$,~$v$)-distance spacings had been removed.

The synthesized beam widths after continuum self-cal were changed by factors of 1.0, 1.6, and 3.7 in Bands 8--10, respectively. We therefore took the data for each band with only B2B phase referencing applied and flagged the same antennas/scans as were lost due to the failed solutions in continuum self-cal. This enabled us to make target images at the same resolution as the continuum self-calibrated images, in order to compare the recovered flux and assess the image coherence with B2B phase referencing alone. 

\subsubsection{
  Band~10 HCN maser self-cal
}\label{sec:03-05-02}

Figure~\ref{fig:04}
shows the detected HCN maser cross-correlated spectrum for a pair of antennas with the shortest projected baseline length of 5259$-$5290~k$\lambda$. As is indicated from the line profile, creating the HCN maser cube confirmed that the brightest HCN maser emission was located in the spectral channel at the LSRK velocity of 9.4~km~s$^{-1}$ and that the peak emission region was spatially compact. The image model of this single channel was used for phase self-cal with the solution interval of $\sim 6$~s (approximately the correlator data integration period). This produced an improved HCN maser image at the LSRK velocity of 9.4~km~s$^{-1}$, and this model was then used for amplitude self-cal with the solution interval of the scan length (25--30~s). These phase and amplitude self-cal corrections were then applied to the entire Band~10 HCN maser as well as the continuum data sets using {\sc CASA} {\tt applycal} with {\tt interp=`linearPD'} to each integration and scan without interpolating in time. This allowed imaging of the self-cal Band~10 R~Lep HCN maser and the continuum emission at the full angular resolution.

\section{
  Results
}\label{sec:04}
\subsection{
  Image coherence using the continuum self-cal
}\label{sec:04-01}

The images of the continuum emission of R~Lep with B2B phase referencing alone are shown in the left panels of 
Figure~\ref{fig:05} 
in Bands~8 and 9, and the left panel of 
Figure~\ref{fig:06} 
in Band~10. 
Table~\ref{tbl:07} 
summarizes image characteristics of the continuum images with B2B phase referencing alone including the beam size, image peak flux density, image rms noise, and the peak brightness temperature. These images are at the highest possible angular resolution in each HF band. The Band 10 image has a synthesized beam of $4.9 \times 4.7$~mas, which is the maximum angular resolution that ALMA is capable of. 

The images with B2B phase referencing and continuum self-cal (denoted, B2B phase referencing$+$continuum self-cal) are shown in the right panels of 
Figure~\ref{fig:05} 
in Bands~8 and 9, and the middle panel of 
Figure~\ref{fig:06} 
in Band~10. 
Table~\ref{tbl:08} 
summarizes the characteristics of the B2B phase referencing$+$continuum self-cal images. Because the longer baselines were flagged out especially in Bands~9 and 10 as explained in 
Section~\ref{sec:03-05-01}, 
the beam size of the B2B phase referencing$+$continuum self-cal became wider than that with B2B phase referencing alone. 
To align the beam size in each band and make the comparisons of the images between B2B phase referencing alone and B2B phase referencing$+$continuum self-cal, we made images with only B2B phase referencing applied but also the flags that occurred during the continuum self-cal that accordingly removed the longest ($u$,~$v$)-distance spacings. The image coherence with B2B phase referencing alone but with the continuum self-cal flags in Bands~8--10 is 92\%, 83\%, and 77\%, respectively, as listed in the last column of 
Table~\ref{tbl:08}. 
The obtained image coherence for the continuum emission would be a representative value for array configurations with $B_{\mathrm{max}} \leq 16$~km, for Bands~9 and 10, while in Band~8 we obtained the high image coherence of 92\% for the full 16~km scale array configuration. Considering the beam size of the B2B phase referencing$+$self-cal images, the effective longest baseline for the R~Lep continuum source when doing continuum self-cal is $\sim 10$ and $\sim 5$~km in Bands~9 and 10, respectively.

\subsection{
  Image coherence in Band~10 using the HCN maser self-cal
}\label{sec:04-02}

The image characteristics of the HCN maser at the LSRK velocity of $9.4$~km~s$^{-1}$ are listed in 
Table~\ref{tbl:07}, 
and those of the HCN maser at the same velocity channel and the Band~10 continuum calibrated with B2B phase referencing$+$HCN maser self-cal are listed in 
Table~\ref{tbl:09}. 
The right panel of 
Figure~\ref{fig:06} 
shows the Band~10 continuum image calibrated with B2B phase referencing$+$HCN maser self-cal. 
Figure~\ref{fig:07} 
shows the resultant cube made with B2B phase referencing$+$HCN maser self-cal by averaging every four contiguous velocity channels in imaging, corresponding to a velocity width of 1.3~km~s$^{-1}$. The averaged HCN maser emission in the velocity range between $-4.4$ and $21.6$~km~s$^{-1}$ is shown in 
Figure~\ref{fig:08}. 

As mentioned in 
Section~\ref{sec:04-01}, 
we made images with only B2B phase referencing applied but also the HCN maser self-cal flags for the HCN maser at the LSRK velocity of 9.4~km~s$^{-1}$ and the Band~10 continuum. Because we could successfully obtain the self-cal solutions for the full ($u$,~$v$) coverage, thanks to the compact and bright emission of the HCN maser, the beam size had little change after the HCN maser self-cal. 
The image peak flux density of the HCN maser at the LSRK velocity of $9.4$~km~s$^{-1}$ changed from 735~Jy 
(Table~\ref{tbl:07} 
with B2B phase referencing alone) to 1247~Jy by self-cal afterward 
(Table~\ref{tbl:09}). 
Thus, the image coherence at the brightest HCN maser velocity channel is 61\% with B2B phase referencing alone. 
For the Band~10 continuum emission, we find that the image coherence achieved using B2B phase referencing was 70\%. 
The image rms noise of the Band~10 continuum slightly increased from 0.8 to 1.0~mJy~beam$^{-1}$ after applying the flags of the HCN maser self-cal. We attribute this to the reduction of the number of visibilities. After applying B2B phase referencing$+$HCN maser self-cal, the image rms noise is reduced down to 0.9~mJy~beam$^{-1}$, improving the image S/N from 29 to 40.

The image coherence of the Band~10 continuum image calibrated with B2B phase referencing alone of 70\% is also larger than the 61\% achieved on the single channel for the HCN maser. We attribute this to the fact that the continuum emission is centrally focused, whereas the HCN maser emission forms a ring structure wherein the distribution of flux distribution changes slightly before and after self-cal, i.e., we are applying a rule to measure image coherence for pointlike sources to something that has an extended distribution. 

\subsection{
  Effectiveness of continuum self-cal for R~Lep
}\label{sec:04-03}

\citet[][Section~4.4.2]{2022arXiv220705591R} 
describes an analytical relation in self-cal between the solution interval, the number of antennas, and S/N. Applying their analysis to the initial S/N of 29 of the Band~10 continuum image, the per antenna, per scan S/N is $<$2 for 43 antennas and 50 scans, so it is not surprising that not all antenna solutions succeeded (even if considering the ideal peak flux after applying the B2B phase referencing$+$HCN maser self-cal to the Band~10 continuum, 36.0~mJy~beam$^{-1}$, the S/N per antenna, per scan remains $<$2). The degraded angular resolution (wider synthesized beam) image obtained after continuum self-cal yields an image S/N $>$100 and thus provides an S/N per antenna, per scan of $>$3 for the number of antennas not flagged by failed solutions 
(Table~\ref{tbl:08}). 
Hence, limiting the minimum S/N for the self-cal solutions is obeyed, and solutions on antennas below this are flagged.  
The Band~9 continuum S/N per antenna, per scan, is also slightly $<$3 at the start of self-cal, and hence we again lose longer baseline data, although not as many as for Band~10. We do not pursue low-S/N mitigation strategies for self-cal of the continuum such as described in, 
e.g., \citet[][Section~5.10]{2022arXiv220705591R}. We note that if using lower S/N limits for self-cal solutions it might have been possible to save some longer ($u$,~$v$)-distance spacings at the expense of solution accuracy.

\subsection{
  Interpretation of the R~Lep images
}\label{sec:04-04}

The uniform disk fitting analysis shows the diameter of the R~Lep continuum source is $\sim 14-18$~mas as listed in 
Table~\ref{tbl:06}, 
corresponding to 3$-$4~au submillimeter stellar radius at the distance of R~Lep. 
\cite{2019A&A...626A..81V} 
found that O-rich AGB stars have mm-wave radio atmospheres 15\%--50\% larger than the optical/IR stellar radius. C-rich stars have been less studied except for IRC+10216, where  
\cite{2018AJ....156...15M}  
suggest a similar relationship. 
Our measured diameter at 0.33~mm wavelength is 
$\sim 20$\% larger than the IR diameter 
\citep{2005ESASP.560..651H}, 
consistent with expectations.

At the angular resolution of 5~mas (corresponding to a 2.4~au linear scale at 
the distance of R~Lep) 
we can characterize the location of the HCN maser 
distribution as a ring-like morphology with a diameter of 10$-$60~au, 
surrounding the Band~10 submillimeter-wave continuum emission.
The extent of the HCN maser regions was found to be $\sim$10$-$30~au in V~Hya and IRC$+$10216 \citep{Wong2019}, and our imaging result for the HCN maser in R~Lep supports those previous works. 
We confirm the previous prediction that the maser emission must originate from the innermost regions of the CSE due to the high energy level ($E_{\rm low}/k>4200$ K) of the HCN maser line 
\citep{2003ApJ...583..446S}.
This is also consistent with the observations of other HCN maser lines, which indicate that they arise in the wind-acceleration regions of carbon-rich AGB stars 
\citep[e.g.,][]{Lucas1989,Menten2018,2022A&A...666A..69J}. 
Scientifically, the high angular resolution images provided by these technical tests can be used to characterize the inner wind of R~Lep and facilitate the study of different processes close to the star, such as shocks driven by stellar pulsation, dust formation, and wind acceleration
\citep{Hoefner2018},
in unprecedented detail.

\section{
  Discussion
}\label{sec:05}
\subsection{
  Assessment of the goals of the experiments
}\label{sec:05-01}

Here we review the overarching goal of our experiments as described in 
Section~\ref{sec:02-02}, 
as to achieve an image coherence of $\geq 70$\%. 

\subsubsection{
  B2B phase referencing 
}\label{sec:05-01-01}

According to the ALMA calibrator source catalog, the phase calibrator J0504-1512 had the flux density of 0.15 and 0.10~Jy in Bands~3 and 7, respectively, in 2019 September, the most recent measurement at the time of our experiments. Assuming that the flux density is proportional to $\nu^{\alpha}$, the extrapolated flux density to Bands~8--10 is 0.09, 0.08, and 0.07~Jy, respectively, with the constant $\alpha$ of $-0.36$. If we assume an in-band phase referencing scenario, using 43 12~m antennas with four, eight, and eight 2~GHz bandwidth spectral windows in Bands~8--10, respectively, then even when combining the available spectral windows we can only achieve the S/Ns of 14, 6, and 3 in Bands~8--10, respectively, for a scan length of 13~s. 
Since the required S/N for a single phase 
calibrator scan is 15 as to provide qualified phase solutions, 
J0504-1512 is thus too weak to be detected with sufficient S/N 
in the short scan duration required for the fast switching phase calibration 
in Bands~9 and 10, 
while it could plausibly be used in Band~8. 
With in-band phase referencing we would have to 
search for a quasar brighter than 572~mJy in Band~10 
\citepalias{A20}. 
The nearest quasar bright enough is J0522-3627 
at the distant separation angle of $22^{\circ}$, in violation of the requirements stated by \citet{2023ApJS..267...24M}, and if used, it would mean that R~Lep would not have been well calibrated 
(see Section~\ref{sec:05-03}). 
Hence, B2B phase referencing provides us with the only way to use J0504-1512 as the phase calibrator for Band~9 and 10 observations. 

\subsubsection{
  Antenna fast switching and experiment Go/NoGo 
}\label{sec:05-01-02}

The R~Lep continuum images with B2B phase referencing alone compared to those with self-cal show that the image coherence in Bands~8--10 is 92\%, 83\%, and 77\%, respectively (although with a degraded angular resolution for Bands 9 and 10, except when using the HCN maser self-cal where Band~10 achieved 70\%). Meanwhile, the modified coherence factors obtained from the Go/NoGo checks 
(Equation~(\ref{eq:02})) 
were 95\%, 85\%, and 77\%, respectively. Technically, the Go/NoGo check provides a prediction of the phase stability of a single source, hence providing the image coherence for in-band phase referencing with a $0^{\circ}$ phase calibrator separation angle. However, a visibility phase error approximately proportional to the separation angle occurs in real observations 
\citep{2023ApJS..267...24M}. 
In addition, [\citetalias{M20}] 
report that the image coherence in B2B phase referencing tends to be lower than that in in-band phase referencing because of the DGC process where a minor phase offset error can be propagated to the target visibility data. Although the Go/NoGo check cannot estimate the phase error propagation, the modified coherence factor is a reasonably good indicator in predicting the image coherence of these HF long baseline observations. 
 
In the process of our experiments, we learned two important lessons related to achieving the required image coherence. First is the Go/NoGo phase rms condition. The phase rms of $63^{\circ}$ in the B-10 Go/NoGo check was less stable than our ``GO" criterion of $45^{\circ}$ and is the very upper limit of what was acceptable. The worse-than-optimal stability conditions may be the main reason for the low image coherence in Band~10 compared to the other two bands. On the other hand, it is rare to find conditions better than our phase rms threshold even at the ALMA site. Recent analysis of the ALMA site weather conditions, in term of phase stability, show us that the fraction of suitable time meeting our low phase rms criteria is on average about 5\%--10\%, but it depends on the season: if we limit only to winter nights, the fraction is $> 10$\% 
\citep{2023arXiv230408318M}. 
The same report also illustrates a higher fraction of time available with shorter switching cycle times, and hence in the future, our Go/NoGo threshold could be relaxed (increase) if we were to shorten $t_{\mathrm{swt}}$, although at the expense of observational efficiency. 

\begin{figure*}
\gridline{\fig{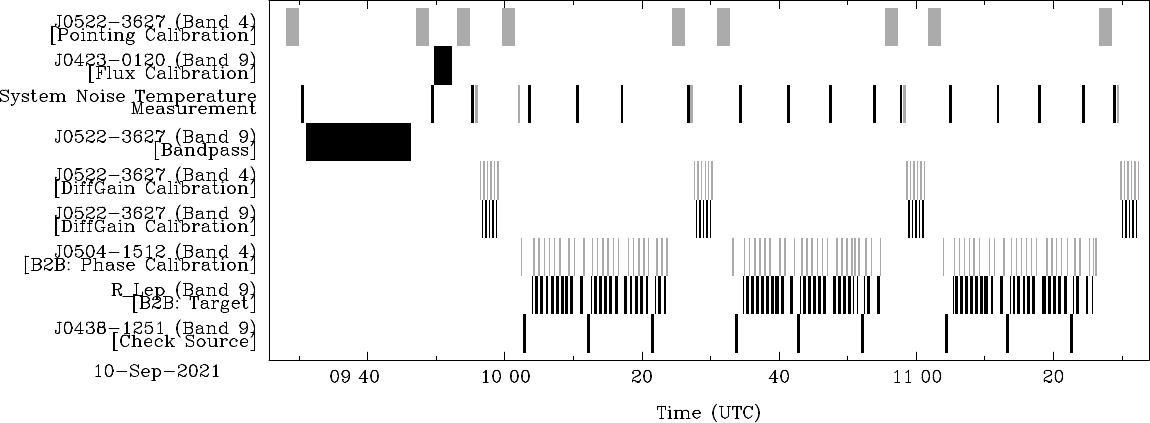}{15cm}{}
          }
\caption{
Experiment schedule of the B-09 B2B phase referencing test. Black bars represent HF band scans in Band~9, while gray bars represent LF band scans in Band~4. The horizontal axis is the experiment date and time in UTC. The observed sources, bands (in parentheses), and scan intents (in square brackets) are found in the left part of the panels. 
\label{fig:01}}
\end{figure*}

\begin{figure*}
\gridline{\fig{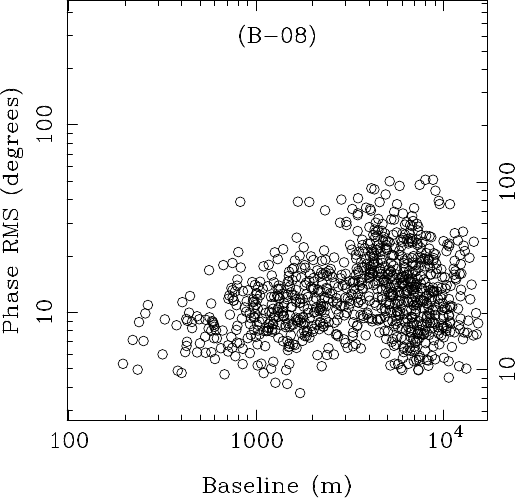}{0.320\textwidth}{}
          \fig{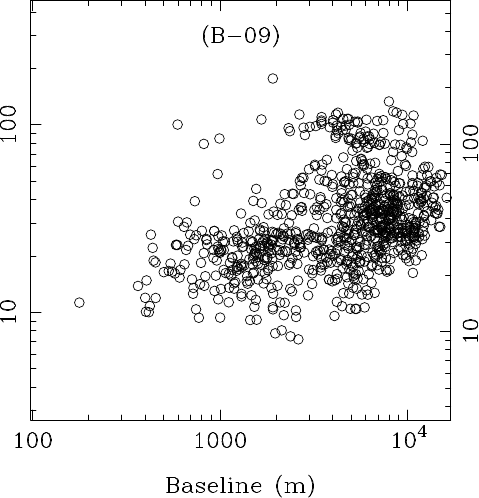}{0.298\textwidth}{}
          \fig{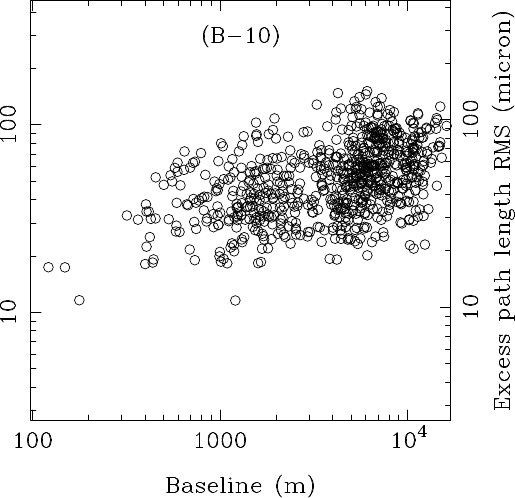}{0.320\textwidth}{}
          }
\caption{
Rms phase as a function of baseline length of the 120~s Go/NoGo checks for J0423-0120 just before the experiments. 
The horizontal axis denotes the baseline length, and each open circle denotes the phase rms value in degrees at $\nu_{\mathrm{_{\mathrm{HF}}}}$ for a single baseline. 
The flux density of J0423-0120 was 
bright (6.4, 3.4, and 2.6~Jy at Bands~3, 7, and 9, respectively) and 
the thermal phase 
noise is negligible after averaging the available spectral windows and 
the dual polarization pairs. 
Note that the corresponding excess path length rms 
is expressed in the second vertical axis in microns at the right side.
\label{fig:02}}
\end{figure*}

Second is the uneven distribution of the switching cycle times. 
Figure~\ref{fig:09} 
shows a histogram of $t_{\mathrm{swt}}$. Typically, $t_{\mathrm{swt}}$ is around 50~s in our experiments, while much longer gaps between phase calibration scans sometimes happened when the check source and/or $T_{\mathrm{sys}}$ measurement scans were inserted, as depicted in 
Figure~\ref{fig:01}. 
Thus, about a quarter to a third of the target scans had longer $t_{\mathrm{swt}}$ than 60~s in the experiments 
(Figure~\ref{fig:09}). 
Those scans were usable, but not ideal, because the longer $t_{\mathrm{swt}}$ means the phase calibration is not as optimal, and hence allows for a larger decoherence. The timing of the insertion of check source and $T_{\mathrm{sys}}$ measurement scans is controlled by the ALMA online software, which, as a result of our experiments, has been improved from Cycle~10 for HF long baseline observations, by putting those extra scans outside of the target and phase calibrator sequence in order not to cause longer $t_{\mathrm{swt}}$. 

\subsection{
  Effect of morphology of the phase calibrator and DGC source 
  on the target images 
}\label{sec:05-02}

In the imaging of the phase calibrator (J0504-1512) and the DGC source (J0522-3627) we found low flux level minor extended structures. 
In general, if a phase calibrator has resolved structure, applying its phase solutions to the target may cause the pattern to be mirrored in the target image, therefore ALMA does have a guideline that any extended flux density or long baseline resolution effects should be less than 10\% of the core flux for the source to be reliable as the DGC, bandpass or phase calibrator. Of course, the source must be sufficiently strong so that the calibration solutions from it are not significantly S/N limited. The difficulty in finding calibrators at high frequencies while using the longest baseline configuration is that there are few sufficiently bright, perfect pointlike quasars available for DGC or phase referencing, and so calibrators that are slightly extended or resolved must be considered. In the case of B2B phase referencing, the systematic phase error due to the calibrator's structure will also be scaled up by multiplying by $R$ (although the angular resolution will also be a factor of $R$ worse, such that there is a lower likelihood of resolving the calibrators) and so we reconducted our reduction to examine any negative effect.

Figure~\ref{fig:10}
shows synthesized images of J0504-1512 at $\nu_{\mathrm{_{\mathrm{LF}}}}$ (left panels), and J0522-3627 at $\nu_{\mathrm{_{\mathrm{LF}}}}$ and $\nu_{\mathrm{_{\mathrm{HF}}}}$(middle and right panels, respectively). These show that calibrator sources have extended morphology. In order to investigate how the calibrator structure affects the target image quality, we recreated the target images with B2B phase referencing alone after the structure models of J0504-1512 and J0522-3627 generated from the CLEAN components were provided to the ``MODEL'' column of the MSs to regenerate the phase and amplitude calibration solutions as described in 
Section~\ref{sec:03}. 

The top panels of 
Figure~\ref{fig:11} 
show the recreated continuum images with B2B 
phase referencing 
by adopting the 
structure models of the phase calibrator and DGC source. 
The difference from the original B2B phase referencing images (assuming calibrators as point sources) as shown in the left panels of 
Figures~\ref{fig:05} 
and 
\ref{fig:06} 
can be hardly recognized. The entire imaged regions are consistent within $5\sigma$ 
(image rms noise) in Bands~8--10 as shown in the bottom panels of 
Figure~\ref{fig:11}, 
which depicts the per-pixel differences normalized by the 
image rms noise of each original target image 
(Figures~\ref{fig:05} 
and 
\ref{fig:06}).
Thus the effect of calibrators' structure is well within the tolerance of the offered ALMA amplitude accuracy of 10\% in Band~8 and 20\% in Bands~9 and 10
\citep{Privon2022} 
for the the R~Lep images with dynamic ranges of 255, 98, and 27 in Bands~8--10, respectively. We note that if a high dynamic range is required to image relatively faint structures in a target source, that is a dynamic range down to 3 to orders of magnitude lower than the peak, then the non-pointlike structure of the calibrators may produce image distortions in the outskirts of the brightness distribution. Here for R~Lep we consider the differences insignificant.

\subsection{
  Image quality of check sources as a function of 
  separation angle 
}\label{sec:05-03}

In each experiment, we observed one additional quasar (a check source) to independently assess the data quality. This is a standard procedure for ALMA HF and long baseline observations. The premise of the check source is to be equally distant from the phase calibrator as the science target is, so that phase solutions from the phase calibrator applied to the check source would be representative of the calibration that the target would receive. However, it is not possible in reality to find such a bright enough quasar in the HF bands, so that the check source is automatically selected in a query to the ALMA online software, which we allowed to select a bright quasar without any restriction on the separation angle to the phase calibrator. 
In our experiments, 
J0516-1603, 
J0438-1251, and 
J0423-0120 
were selected 
in B-08--B-10, respectively, as listed in 
Table~\ref{tbl:02}. 
The separation angle 
between the check source and phase calibrator (J0504$-$1512) is 
$3^{\circ}.0$, $6^{\circ}.7$, and $17^{\circ}.2$ in 
B-08--B-10, respectively. Already based upon our previous studies 
(\citetalias{M20}; \cite{2023ApJS..267...24M}),
the separation angle for the B-09 and B-10 experiments is too large to be fully representative of the calibration made to R~Lep using a phase calibrator at only $1^\circ.2$ separation. 

Figure~\ref{fig:12} 
shows the synthesized images of the check sources of J0516-1603, J0438-1251, and J0423-0120 in Bands 8--10, respectively. The detection in Bands~8 and 9 can be confirmed, Band 8 appearing pointlike indicative of good phase calibration, while that in Band 9 is distorted. In Band~10, there is no detection of the known point source. This illustrates the combined effects of increasing frequency and increasing angular separation on the reducing the effectiveness of phase referencing. This reinforces the need to use a phase calibrator at a closer angular separation in general, and moreover highlights the use of B2B phase referencing as the mode to provide a phase calibrator within $< 3^{\circ}$ in the HF bands. As noted, check sources should be found equidistant from the phase calibrator as the target source is, although given the low likelihood of finding a phase calibrator in the first instance, the possibility of finding another quasar at the HF bands is very difficult. The images of check sources that are further away from the phase calibrator than the target must be interpreted with caution. 

\subsection{
  User's perspective for HF-LB observations in ALMA Cycle~9
}\label{sec:05-04}

From the user perspective, there is no required special treatment for HF long baseline observations as are offered from 2022 October in Cycle~9. The user can set up their proposals in the ALMA Observing Tool (OT) in the same fashion as any other frequency band or configuration project. In particular, based upon the aforementioned HF-LBCs, the work detailed in \citet{2023ApJS..267...24M} and in this paper, the OT checks the ALMA calibrator source catalog for suitable bright and close point-source phase calibrators during the project validation for Bands 8--10 observations with long baseline configurations. Information about the stringent angular separation limits for the phase calibrator to target separation are detailed in the ALMA Proposer's Guide \citep{Privon2022}. 

Due to the difficulty in finding bright enough phase calibrators at the high frequencies, it is a recommended practice for users to include wider bandwidth spectral windows in order to raise the gain solution S/N if such a setup is feasible for their given science proposal. If the OT search cannot find a suitable phase calibrator, it automatically switches to a B2B phase referencing calibrator and searches at a predetermined lower frequency band (\citetalias{A20}). The observation feasibility for approved proposals is ultimately confirmed by the ALMA observatory prior to the observation taking place. During any B2B phase referencing mode observations the DGC source is added into the observing run by the ALMA online software, as is done for standard calibrators such as the bandpass calibrator in normal observations.

\section{
  Summary
}\label{sec:06}

The HF-LBC-2021 experiments to image the carbon-rich star R~Lep demonstrated that the highest angular resolution ALMA observation at the largest configuration and highest frequencies can produce images with an image coherence of $>$70\%.  The use of B2B phase referencing, especially in Bands~9 and 10, will normally be needed to find a suitable phase calibrator close enough to the target.
In order to reach the image coherence of $\geq 70$\%, 
observations must be made under stable atmospheric conditions as  judged from 
the 120~s Go/NoGo check before an intended observation. A reasonable Go/NoGo criterion is that
the 120~s phase rms is $\leq 63^{\circ}$ for the longest quartile of the baseline lengths, although a value of $\leq 45^{\circ}$ is preferred and will result in improved image coherence. This Go/NoGo criterion applies to ALMA HF long baseline observations from Cycle~9 (2022 October). Fast switching between the phase calibrator and the target source during phase referencing also allows for good phase calibration, correcting the majority of the fast-changing tropospheric phase errors above each antenna. 

The images of the ALMA online software selected check sources highlighted how increasing distance from the phase calibrator can result in poor calibration. 
The smaller the separation angle from the phase calibrator is, the better the image quality is. The image quality of the check sources demonstrates that the phase calibration can work effectively using nearby phase calibrators of $< 3^{\circ}$ from the target. 
In reality, the phase calibrator should ideally be located within $\sim 1^{\circ}$ of the science target in the highest frequency bands (Bands 9 and 10) with the longest 
baselines of 16~km 
\citep{2023ApJS..267...24M}. The image coherence of R~Lep in all our experiments is $>$70\%, achievable by following the Go/NoGo criteria and using a close phase calibrator.
B2B phase referencing is likely the only method to find sufficiently close phase calibrators as to achieve high phase calibration performance.

The submillimeter photosphere and HCN maser CSE ring of R~Lep were 
imaged with a 5~mas synthesized beam in Band~10.
Self-cal did improve the image quality for the 
bright and compact HCN maser channel at the LSRK velocity of 9.4~km~s$^{-1}$ which was then transferred to the other HCN maser channels and to the Band~10 continuum data.
However, self-cal is not always applicable to all science targets, 
especially with low S/N. Here for R~Lep, Band~9 and 10 continuum self-cal resulted in the loss of the longest baselines and demonstrates the limit of useful {\it weak} source self-cal. 

Our work has clarified the optimal observation parameters for 
HF long baseline observations with B2B phase referencing as detailed in our previous studies 
(\citetalias{M20}, 
\citetalias{M22},
\citealt{2023ApJS..267...24M}). The knowledge from this work and the validation experiments constituting to the HF-LBCs (\citetalias{A20}, \citealt{2020AJ....160...59A}, \citetalias{M20}, \citetalias{M22}, \citealt{2023ApJS..267...24M})
have been used to enable HF long baseline observations for the scientific community, and the practical lessons learned will also be used to guide future ALMA users' observations at the high frequencies using B2B phase referencing with more compact array configurations.

\begin{acknowledgments}
For this research, we made use of the following ALMA data ADS/JAO.ALMA\#2011.0.00009.E, ADS/JAO.ALMA\#2011.0.00001.CAL. 
ALMA is a partnership of ESO (representing its member states), NSF (USA), NINS (Japan), together with the NRC (Canada), NSC and ASIAA (Taiwan), and KASI (Republic of Korea), in cooperation with the Republic of Chile. The Joint ALMA Observatory is operated by the ESO, AUI/NRAO, and NAOJ. The National Radio Astronomy Observatory is a facility of the National Science Foundation operated under cooperative agreement by Associated Universities, Inc.
The authors thank the Joint ALMA Observatory staff in Chile for performing the challenging HF-LBC-2021 successfully.
With this manuscript being published $\sim$10 years since ALMA's first observations, we are now illustrating the successful use of Band 10, at the highest frequency and with longest baselines possible to achieve the highest angular resolution that ALMA can reach. We acknowledge the efforts of many staff, past and present, who have worked on the high frequency, long baseline, and/or B2B modes over those last 10 years:
Denis A. Barkats, Loreto Barcos-Mu\~{n}oz, Craige Bevil, Andy Biggs, Crystal Brogan, John M. Carpenter, Arancha Castro-Carrizo, Stuartt Corder, Juan Cortes, Paulo Cortes, Richard Hills, Todd Hunter, Violette Impellizzeri, Katharine Johnston,  Seiji Kameno, Ryohei Kawabe, Tim Van Kempen, Cristian Lopez, Robert Lucas, Sergio Martin, Satoki Matshushita, Jennifer Donovan Meyer, Anna Miotello, Koh-Ichiro Morita, José Luis Ortiz, Alison Peck, Neil M. Phillips, Vincent Pietu, Adele Plunkett, Alejandro Saez, Tsuyoshi Sawada, Kimberly Scott, Dick Sramek, Ignacio Toledo, Takafumi Tsukui, Baltasar VIla Vilar\'{o}, Eric Villard, Catherine Vlahakis, Nick Whyborn, Christine Wilson.
We apologize as the list is not exhaustive, given many
staff who work(ed) behind the scenes, on operations, the hardware or software systems and who work diligently without mention in previous documents or reports.
The authors thank an anonymous referee for careful reviews and providing valuable suggestions that greatly improved this paper. 
K.T.W. acknowledges support from the European Research Council (ERC) under the European Union’s Horizon 2020 research and innovation program (grant agreement No.~883867, project EXWINGS).
\end{acknowledgments}

%

\vspace{5mm}
\facilities{ALMA(12~m array)}


\software{CASA \citep{2022PASP..134k4501C}  
          }





\bibliography{hflb2021}{}
\bibliographystyle{aasjournal}



\clearpage
\newpage

%
%

\clearpage
\newpage

\clearpage
\newpage

\begin{figure}[ht!]
\gridline{
    \fig{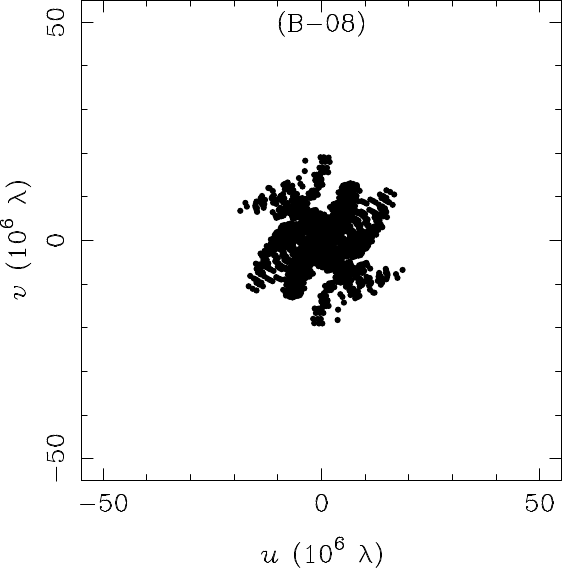}{0.3\textwidth}{}
    \fig{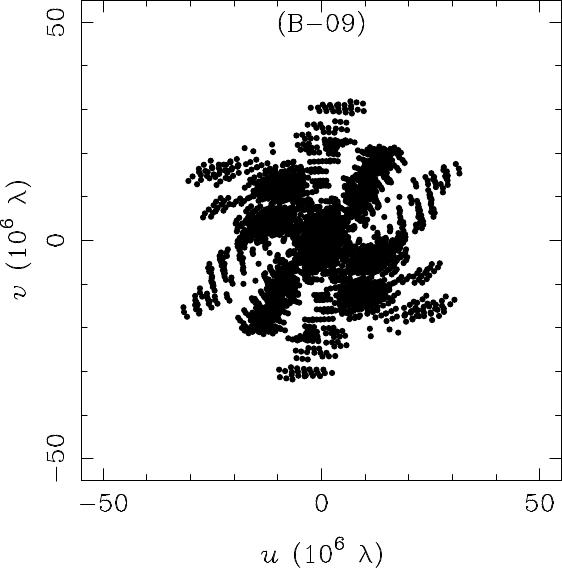}{0.3\textwidth}{}
    \fig{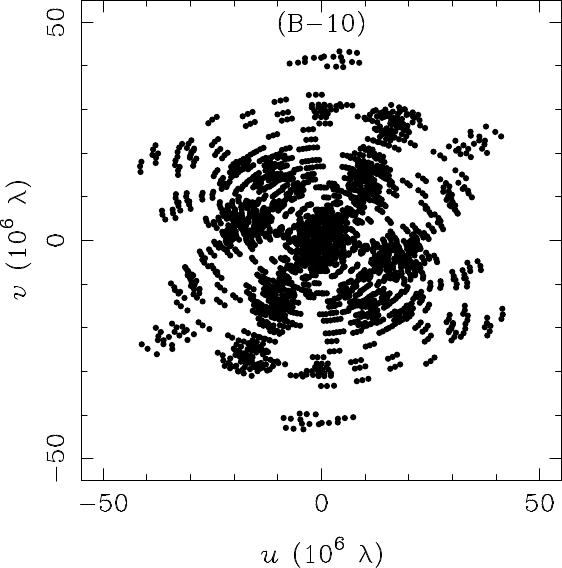}{0.3\textwidth}{}
}
\gridline{
    \fig{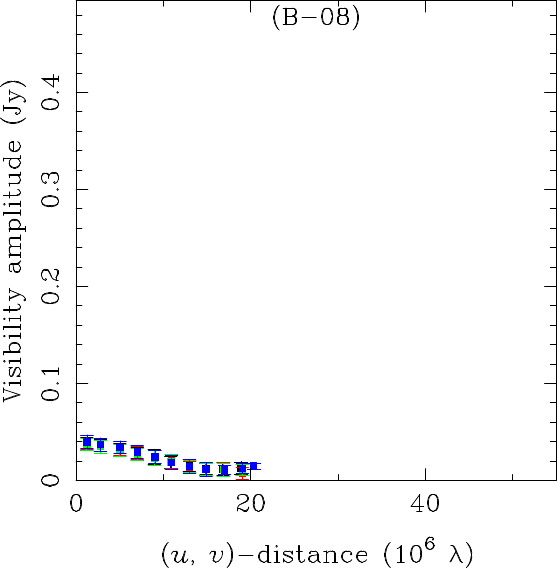}{0.3\textwidth}{}
    \fig{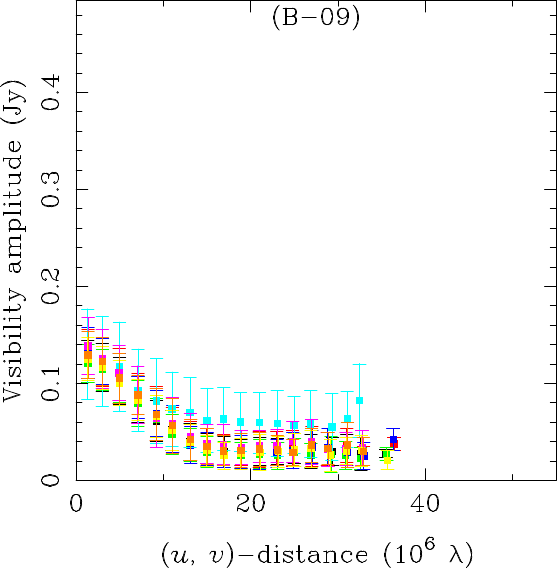}{0.3\textwidth}{}
    \fig{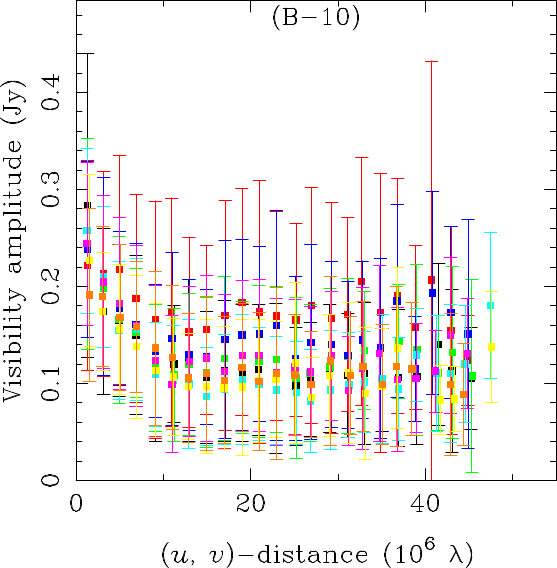}{0.3\textwidth}{}
}
\caption{
Top: ({\it u},~{\it v}) coverage of R~Lep in B-08--B-10 from the left, middle, and right, respectively. 
Bottom: visibility amplitude of the continuum as a function of ($u$,~$v$) distance of R~Lep in B-08--B-10 from the left, middle, and right, respectively. The visibility data were first vector averaged in time and two polarization pairs. Then the data are scalar averaged in each 2~M$\lambda$ bin. The color represents each spectral window. In Band~9, one of the spectral windows in BB-3 (LSB: light blue color) has a systematically higher amplitude and deviation compared to the other spectral windows because a cluster of atmospheric O$_{3}$ absorption lines exists 
\citep{2022A&A...664A.153P} 
in the bandwidth.
\label{}}
\end{figure}

\clearpage
\newpage

\begin{figure}[ht!]
\gridline{
    \fig{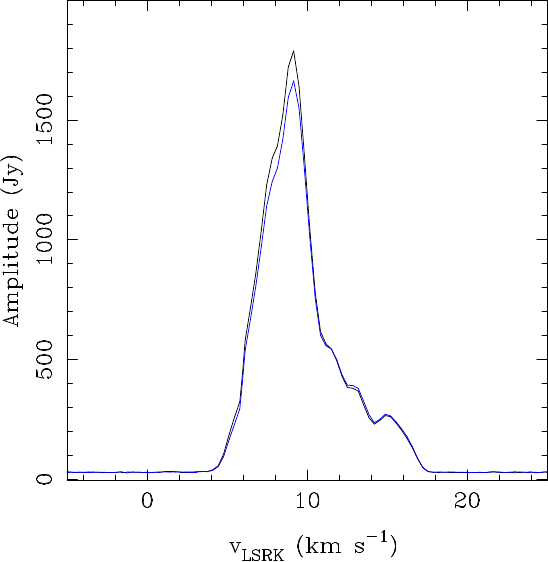}{0.5\textwidth}
    {(B-10: B2B phase referencing$+$HCN maser self-cal)}}
\caption{
Cross-correlation spectrum for the HCN maser 
at 890.8~GHz, from the $J=10-9$ transition 
between the ($11^{1}0$) and ($04^{0}0$) vibrationally excited states, for the shortest projected baseline antenna pair with 
the ({\it u},~{\it v}) distance of 5259--5290~k$\lambda$. 
The horizontal axis represents the radial velocity in 
$v_{\mathrm{_{\mathrm{LSRK}}}}$ (km~s$^{-1}$), and the 
vertical axis represents the cross-correlated flux density (jansky). 
The black and blue lines represent $XX$ and $YY$ polarization pairs, 
respectively. The line profile was obtained by scalar averaging in time. 
\label{fig:04}}
\end{figure}

\clearpage
\newpage

\begin{figure*}
\gridline{
    \fig{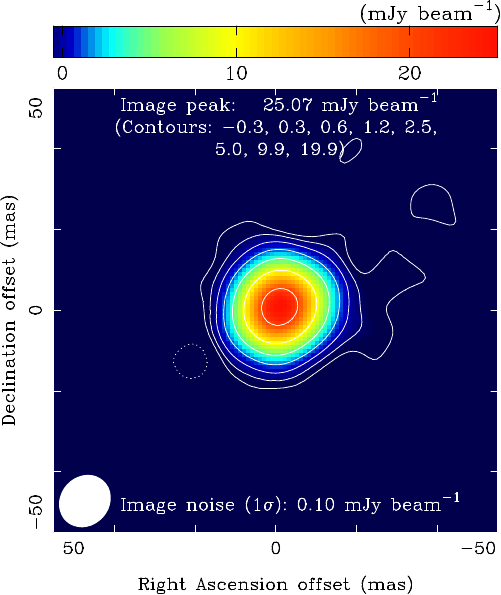}{0.3\textwidth}
    {(B-08: B2B phase referencing alone)
    }
    \fig{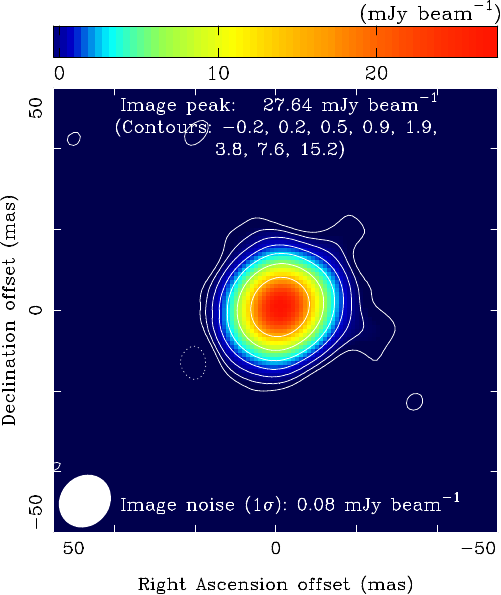}{0.3\textwidth}
    {(B-08: B2B phase referencing$+$continuum self-cal)}
}
\gridline{
    \fig{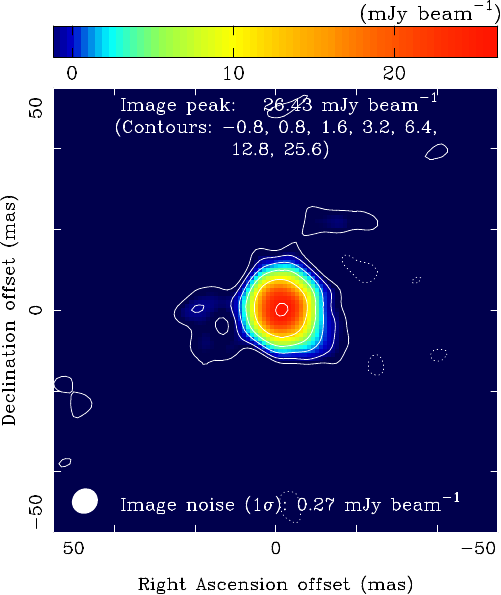}{0.3\textwidth}
    {(B-09: B2B phase referencing alone)
    }
    \fig{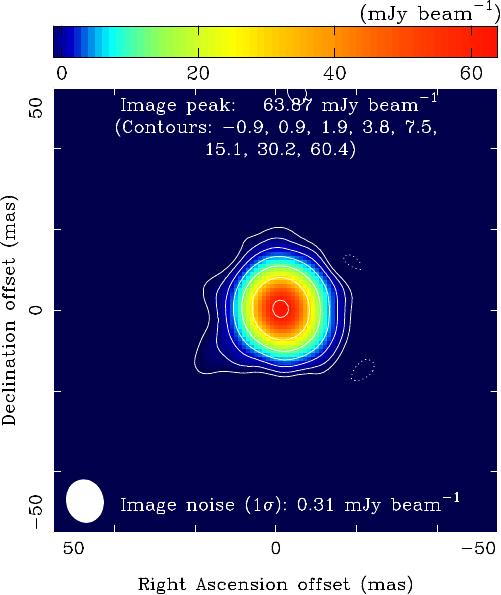}{0.3\textwidth}
    {(B-09: B2B phase referencing$+$continuum self-cal)}
}
\caption{
R~Lep HF long baseline synthesized images for the continuum emission 
at Bands 8 and 9 in the top and bottom, respectively.
The left and right panels are the images before and after phase self-cal (using the continuum), respectively. 
The solid line contours start from $3\sigma$ level, 
increased by a factor of a power of 2, and the dashed line contour represents the $-3\sigma$ level. 
The contour levels are shown on the upper side of each panel. 
The image peak value and rms noise are indicated at the top and bottom of each panel, respectively. 
The ellipse in the lower left represents the synthesized beam.
Note that the synthesized beam size of the continuum self-cal image at Band~9 
becomes broader compared to that of the original images because 
 phase self-cal solutions were not obtained for some of the more distant 
antennas and the data for these were flagged in applying the solutions. 
\label{fig:05}}
\end{figure*}

\clearpage
\newpage

\begin{figure}[ht!]
\gridline{
    \fig{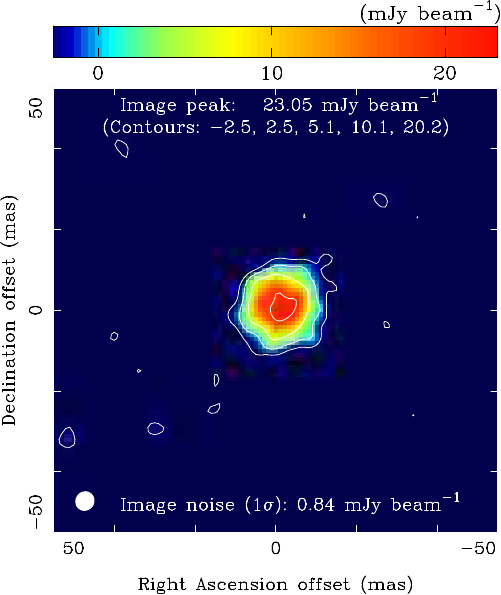}{0.3\textwidth}
    {(B-10: B2B phase referencing alone)
    }
    \fig{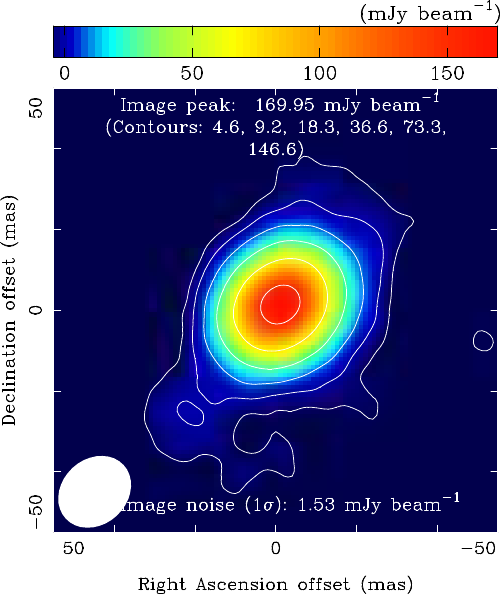}{0.3\textwidth}
    {(B-10: B2B phase referencing$+$continuum self-cal)}
    \fig{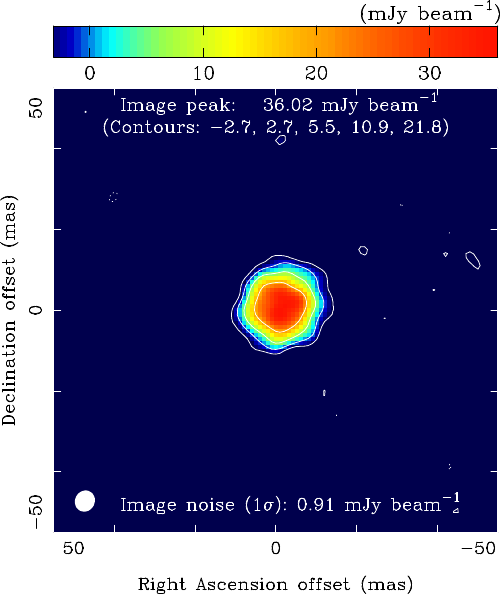}{0.3\textwidth}
    {(B-10: B2B phase referencing$+$HCN maser self-cal)}} 
\caption{
R~Lep Band~10 continuum image. 
Left: B2B phase referencing alone (before phase self-cal).
Middle: after phase self-cal. 
Right: calibrated using the phase and amplitude self-cal 
solutions of the 
HCN maser channel at $v_{\mathrm{LSRK}}=9.4$~km~s$^{-1}$. 
The solid contours start from the $3\sigma$ level, 
increased by a factor of a power of 2, and the dashed line contour denotes the $-3\sigma$ level. 
The contour levels are shown on the upper side of each panel. 
The image peak value and rms noise are listed at the top and bottom of each panel, respectively. The ellipse in the lower left represents the synthesized beam. 
Note that the synthesized beam size of the middle, continuum phase self-cal image  
becomes broader compared to that of the original image (left panel) because 
the phase self-cal solutions were not obtained for some of the more distant 
antennas and the data for these were flagged in applying the solutions. 
\label{fig:06}}
\end{figure}

\clearpage
\newpage

\begin{figure}[ht!]
\plotone{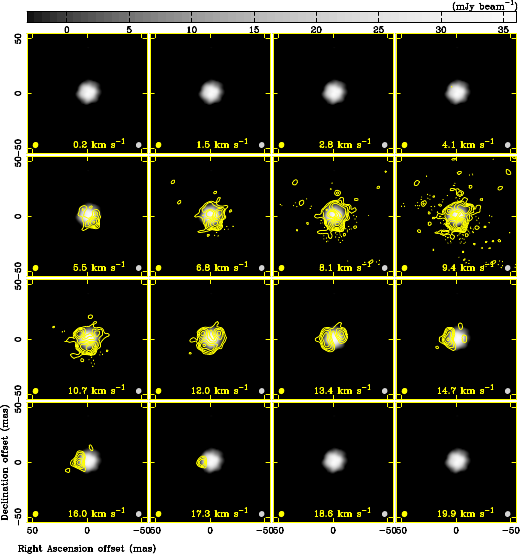}
\caption{
HCN maser cube in Band~10 (yellow contours) 
at 890.8~GHz, from the $J=10-9$ transition 
between the ($11^{1}0$) and ($04^{0}0$) vibrationally excited states, averaging every four velocity channels of the synthesized image cube 
with the velocity width of 0.3~km~s$^{-1}$. The continuum emission in Band~10 is shown as a background grayscale gradation. The image cube was created after applying the amplitude and phase self-cal solutions derived using the HCN maser emission at 
$v_{\mathrm{_{\mathrm{LSRK}}}}=9.4$~km~s$^{-1}$. 
The image rms noise was calculated at each 
velocity channel from emission-free portions 
in the image. 
The minimum image rms noise is 
45~mJy~beam$^{-1}$ 
in the channels with little maser emission at 
$v_{\mathrm{_{\mathrm{LSRK}}}}=19.9$~km~s$^{-1}$,
and the maximum image rms noise is 
1787~mJy~beam$^{-1}$ at $v_{\mathrm{_{\mathrm{LSRK}}}}=9.4$~km~s$^{-1}$. 
The solid line contours start from the $5\sigma$ level per channel at $v_{\mathrm{_{\mathrm{LSRK}}}}=9.4$~km~s$^{-1}$(=8.9~Jy~beam$^{-1}$), 
increased by a factor of a power of 2. The dashed line contours start from the $-5\sigma$ level, increased by a factor of a power of 2. The contour levels are $-17.8$, $-8.9$, 8.9, 17.8, 35.6, 71.2, 142.4, 284.8, and 569.6~mJy~beam$^{-1}$. 
The radial LSRK velocity is given at the bottom of each panel,  
and the synthesized beam size is shown 
on the lower left and right corners for the HCN maser 
($5.6 \times 5.0$~mas)
and 
continuum 
($5.3 \times 4.8$~mas), 
respectively.
\label{fig:07}}
\end{figure}

\clearpage
\newpage

\begin{figure*}
\gridline{
        \fig{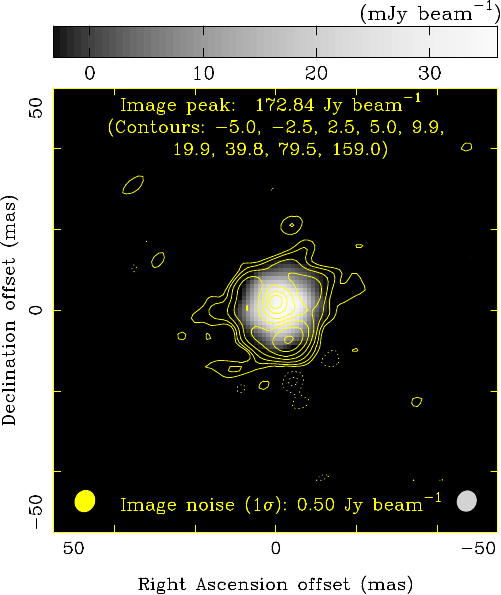}{0.5\textwidth}{}
}
\caption{
R~Lep Band~10 HCN maser 
at 890.8~GHz, from the $J=10-9$ 
transition between the ($11^{1}0$) and ($04^{0}0$) vibrationally 
excited states, 
using the HCN maser 
cube averaged over the velocity range between 
$-0.5$ and 20.3~km~s$^{-1}$ 
in the image cube with the velocity 
width 
of 0.3~km~s$^{-1}$ 
(yellow contours). 
The Band~10 continuum emission is shown as a background grayscale gradation. 
Phase and amplitude self-cal solutions for the HCN maser channel 
at $v_{\mathrm{LSRK}}=9.4$~km~s$^{-1}$ were applied both to the HCN maser 
and the continuum data. 
The solid line contours start from the $5\sigma$ level, 
increased by a factor of a power of 2. 
The dashed line contours start from the $-5\sigma$ level, increased by a factor of a power of 2. 
The contour levels are shown on the upper side. 
The image rms noise of the continuum emission and HCN maser are
0.9 and 
497~mJy~beam$^{-1}$, 
respectively. 
The synthesized beam size is shown 
on the lower left and right corners for the HCN maser 
($5.4 \times 4.9$~mas)
and 
continuum 
($5.3 \times 4.8$~mas), 
respectively.
\label{fig:08}}
\end{figure*}

\clearpage
\newpage

\begin{figure*}
\gridline{\fig{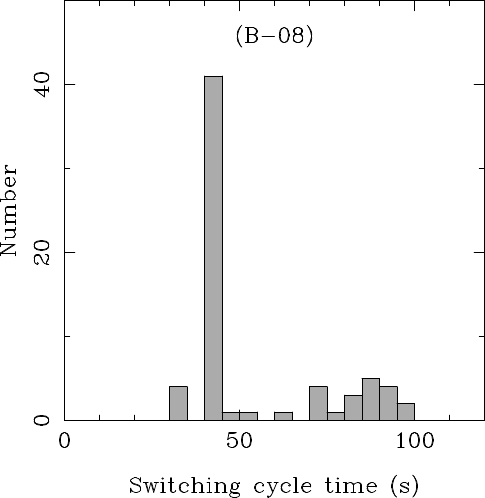}{0.32\textwidth}{}
          \fig{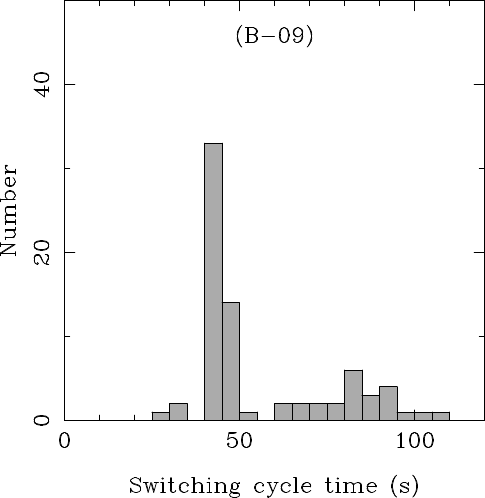}{0.32\textwidth}{}
          \fig{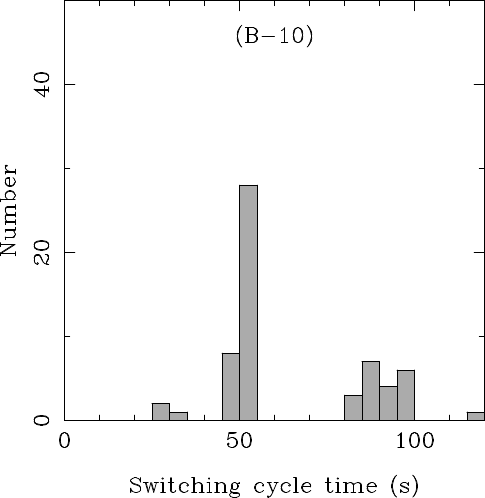}{0.32\textwidth}{}
          }
\caption{
Distribution of the target scans within the switching cycle time 
used during the observations. 
The horizontal axis is the switching cycle time between visits 
to the phase calibrators. 
The longer switching cycle times are due to system noise 
temperature measurement 
or check source scans being inserted in between phase calibrator scans.
From the left, B-08--B-10. 
\label{fig:09}}
\end{figure*}

\clearpage
\newpage

\begin{figure*}
\gridline{\fig{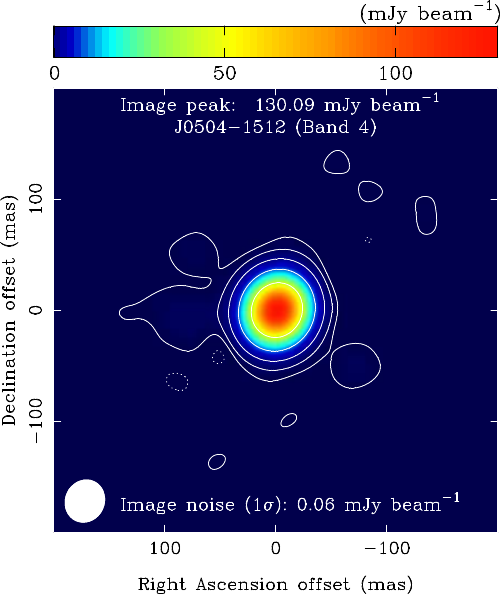}{0.3\textwidth}{(B-08-a)}
          \fig{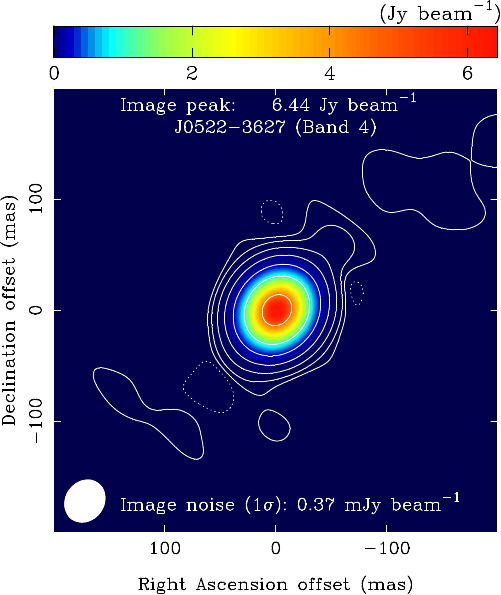}{0.3\textwidth}{(B-08-b)}
          \fig{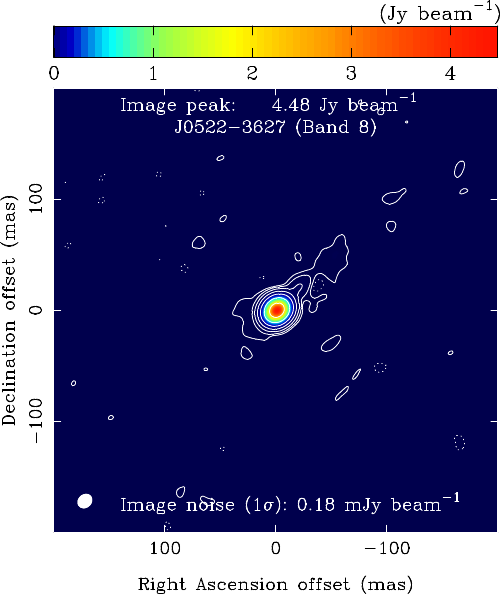}{0.3\textwidth}{(B-08-c)}
          }
\gridline{\fig{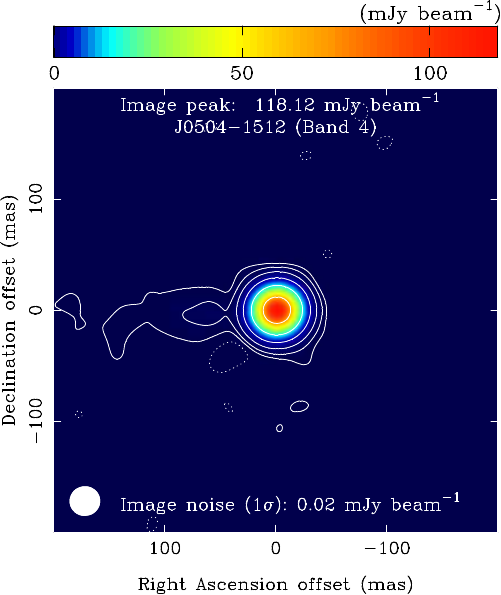}{0.3\textwidth}{(B-09-a)}
          \fig{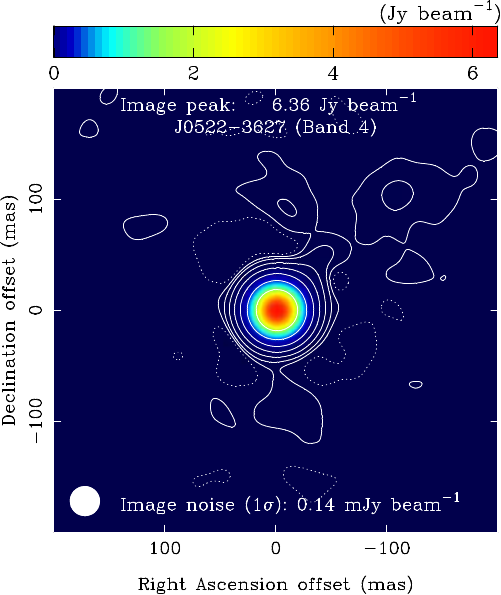}{0.3\textwidth}{(B-09-b)}
          \fig{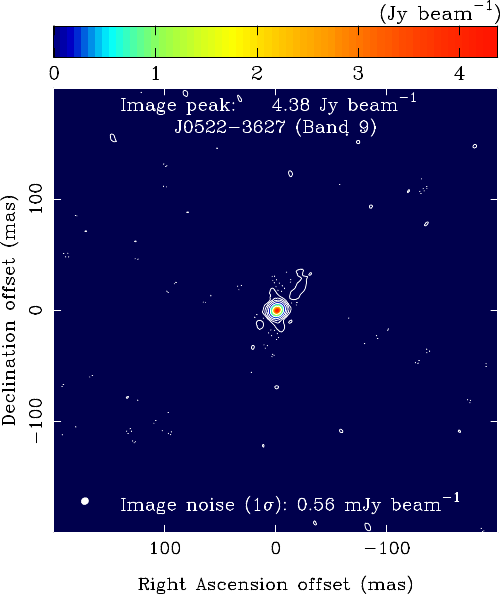}{0.3\textwidth}{(B-09-c)}
          }
\gridline{\fig{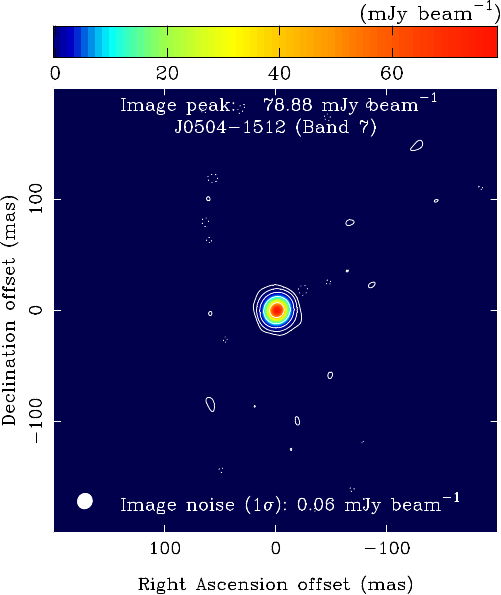}{0.3\textwidth}{(B-10-a)}
          \fig{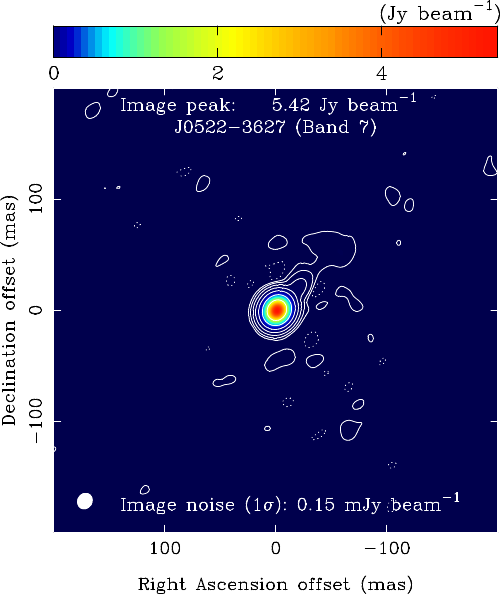}{0.3\textwidth}{(B-10-b)}
          \fig{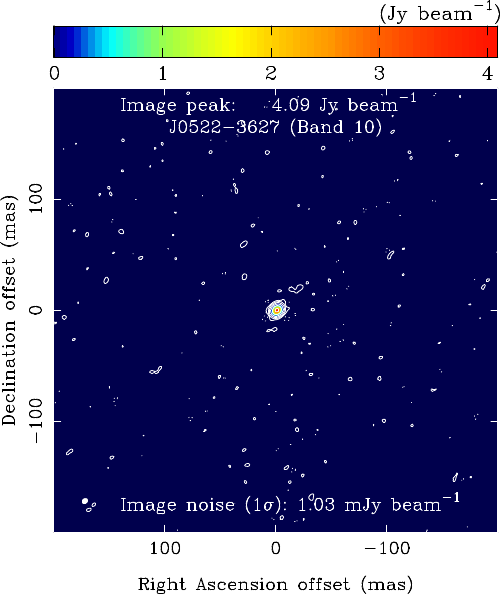}{0.3\textwidth}{(B-10-c)}
          }          
\caption{
\label{fig:10}}
\end{figure*}

\clearpage
\newpage

{\bf Figure~\ref{fig:10}.}
Synthesized images of the calibrators observed in the HF-LBC-2021 experiment.
The top, middle, and bottom rows show the images at B-08--B-10, respectively.
Left, middle, and right panels show the images of  
(a) J0504-1512 (phase calibrator) in the LF band,  
(b) J0522-3627 (DGC source) in the LF band, and 
(c) J0522-3627 (DGC source) in the HF band, respectively. 
The solid line contours start from $3\sigma$ level, increasing by a factor of a power of 4. 
The dashed line contours start from $-3\sigma$ level, increased by a factor of a power of 4. 
The image peak and rms noise are given at the top and bottom of each panel, 
respectively. 
The ellipse in the lower left represents the synthesized beam of the image.

\clearpage
\newpage

\begin{figure*}
\gridline{\fig{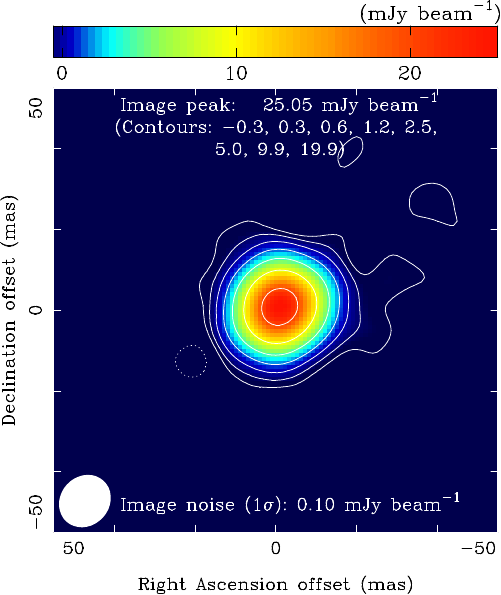}{0.32\textwidth}
              {(B-08 B2B phase referencing alone)}
          \fig{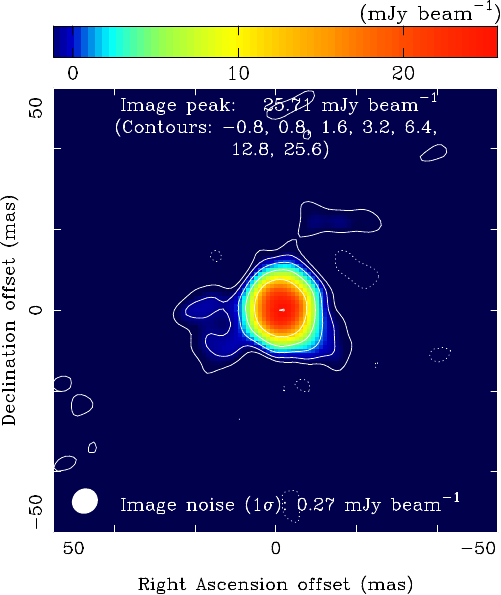}{0.32\textwidth}
              {(B-09 B2B phase referencing alone)}
          \fig{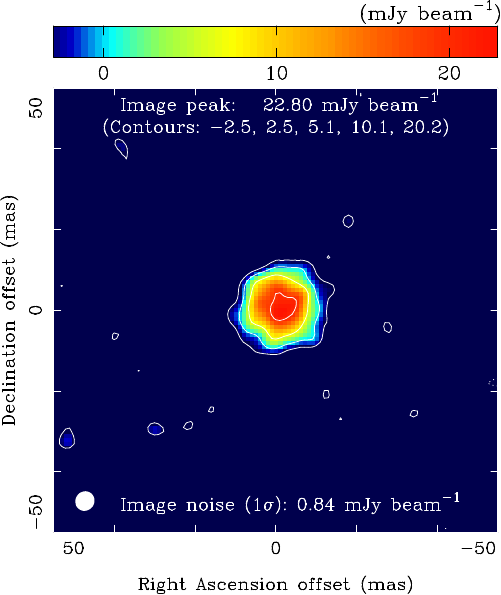}{0.32\textwidth}
              {(B-10 B2B phase referencing alone)}
          }
\gridline{\fig{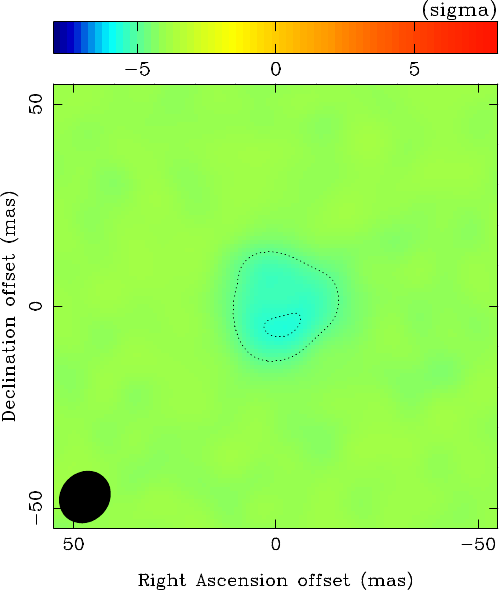}{0.32\textwidth}
              {(B-08 B2B phase referencing alone)}
          \fig{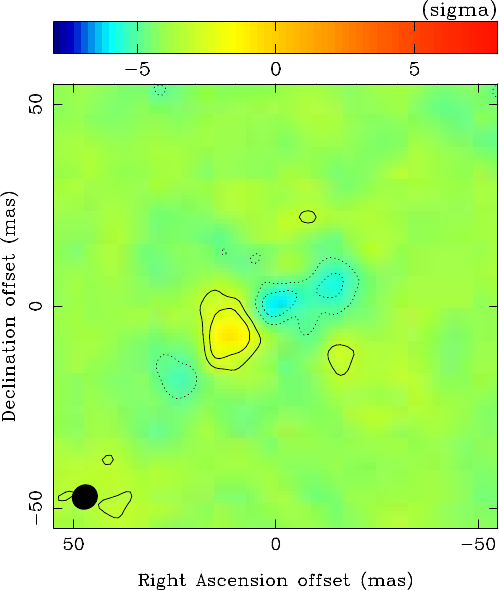}{0.32\textwidth}
              {(B-09 B2B phase referencing alone)}
          \fig{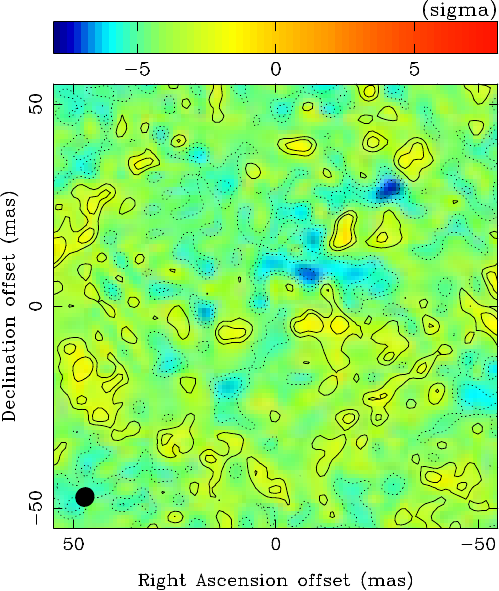}{0.32\textwidth}
              {(B-10 B2B phase referencing alone)}
          }
\caption{
Top: 
R~Lep continuum images calibrated with B2B 
phase referencing 
alone, 
adopting the calibrator morphology model 
as shown in 
Figure~\ref{fig:10}. From the left, B-08--B-10. 
The solid line contours start from the $3\sigma$ level of R~Lep continuum 
image without modeling the calibrator morphology, increased by a factor 
of a power of 2, and the dashed line contour denotes the $-3\sigma$ level. 
The contour levels are shown on the upper side of each panel. 
Bottom: 
differential images of the R~Lep continuum calibrated with 
B2B phase referencing alone between the calibrators' point-source 
assumption and morphology modeling, normalized by the 
image rms noise. 
From the left, B-08--B-10. 
The color scale ranges between $-8 \sigma$ and $8 \sigma$. The contours are 
$-4\sigma$, $-2\sigma$, and $-\sigma$ (dashed lines), and  
$\sigma$, $2\sigma$, and $4\sigma$ 
(solid lines).
%
\label{fig:11}}
\end{figure*}

\clearpage
\newpage

\begin{figure*}
\gridline{\fig{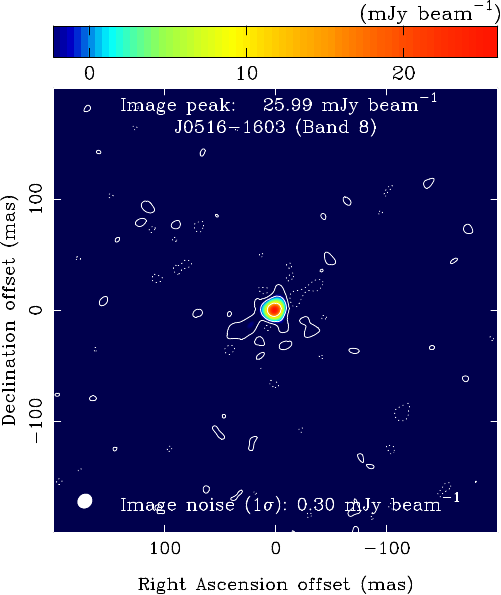}{0.32\textwidth}{(B-08)}
          \fig{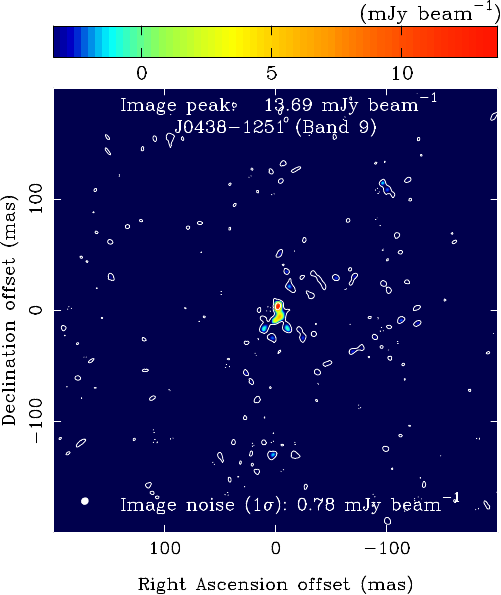}{0.32\textwidth}{(B-09)}
          \fig{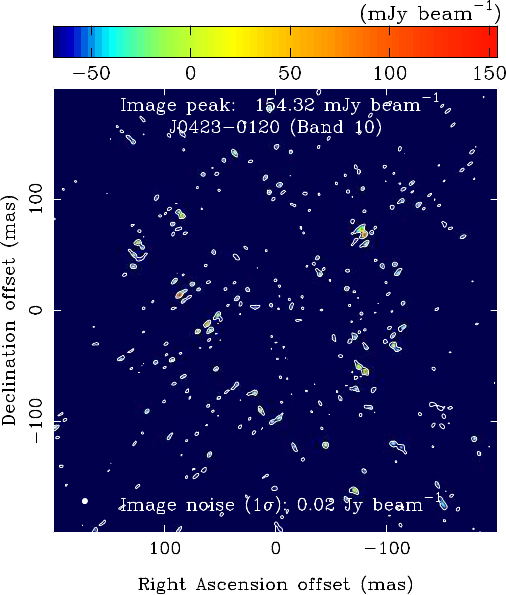}{0.32\textwidth}{(B-10)}
          }
\caption{
Check source images of 
J0516-1603 ($3^{\circ}.0$ from the phase calibrator) in Band~8 (left), 
J0438-1251 ($6^{\circ}.7$) in Band~9 (middle), and 
J0423-0120 ($17^{\circ}.2$) in Band~10 (right).  
The horizontal and vertical axes represent relative R.A. and decl. 
in mas, respectively. The synthesized beam is drawn at the bottom left corner. 
Contours start from the 3-$\sigma$ level, 
increasing by a factor of a power of 4. 
The dashed line contours start from the $-3\sigma$ level, increased by a factor of a power of 4. 
The image peak value and rms noise are listed at the top and bottom 
of each panel, respectively. It is clear that for larger separation angles from the phase calibrator that the check sources cannot be correctly calibrated.
\label{fig:12}}
\end{figure*}

\clearpage
\newpage

%
%
\begin{longrotatetable}
\begin{deluxetable*}{clllllll}
\tablenum{1}
\tablecaption{Experiments in HF-LBC-2021\label{tbl:01}}
\tablewidth{0pt}
\tablehead{
\colhead{Exp Code} 
 & \colhead{Epoch}
 & \colhead{Time (UTC)} 
 & \colhead{HF Band}
 & \colhead{LF Band} 
 & \colhead{Number of} 
 & \colhead{Total ON-source}
 & \colhead{Execution Block}  \\
 &
 & (hh:mm)
 & ($\nu_{\mathrm{_{\mathrm{HF}}}}$ GHz)$^{\mathrm{a}}$ 
 & ($\nu_{\mathrm{_{\mathrm{LF}}}}$ GHz)$^{\mathrm{a}}$  
 & Antennas 
 & Duration for R~Lep (s) 
 & (uid://A002/)
}
\decimalcolnumbers
\startdata
B-08
 & 2021 Sep 23
 & 07:30 $-$ 09:17 
 & Band~8 (405) 
 & Band~4 (135) 
 & 44 
 & 1580 
 & Xf0e992/X4d18\\
B-09
 & 2021 Sep 10
 & 09:28 $-$ 11:33
 & Band~9 (670) 
 & Band~4 (148) 
 & 42 
 & 1643 
 & Xf07268/X32f0\\
B-10
 & 2021 Sep 09
 & 08:44 $-$ 10:20 
 & Band~10 (896) 
 & Band~7 (299) 
 & 39 
 & 1722 
 & Xf06573/X20d3\\
\enddata
\tablecomments{
$^{\mathrm{a}}$ 
Local oscilator frequency of the SIS mixer in the receiver cold cartridge, corresponding to the center RF frequency in the receiving bandwidth. 
}
\end{deluxetable*}
\end{longrotatetable}

\clearpage
\newpage

\begin{deluxetable*}{ccccc}
\tablenum{2}
\tablecaption{Observed Calibrator Sources in the HF-LBC-2021 Experiments\label{tbl:02}}
\tablewidth{0pt}
\tablehead{
\colhead{Exp Code} 
  & \colhead{DGC source}  
  & \colhead{HF Flux Calibrator}  
  & \colhead{HF Bandpass Calibrator}  
  & \colhead{Check Source}  \\
    & 
    & 
    & 
    & \colhead{(Separation from}  \\
      & 
      & 
      & 
      & \colhead{Phase Calibrator)} 
}
\decimalcolnumbers
\startdata
B-08
 & J0522-3627 
 & J0522-3627 
 & J0423-0120 
 & J0516-1603 ($3^{\circ}.0$) \\
B-09
 & J0522-3627 
 & J0522-3627 
 & J0522-3627 
 & J0438-1251 ($6^{\circ}.7$) \\
B-10
 & J0522-3627 
 & J0522-3627 
 & J0522-3627 
 & J0423-0120 ($17^{\circ}.2$) \\
\enddata
\end{deluxetable*}

\begin{deluxetable*}{cccc}
\tablenum{3}
\tablecaption{Spectral Setting of Spectral Windows
\label{tbl:03}}
\tablewidth{0pt}
\tablehead{
\colhead{Base Band No.} 
& \colhead{
Frequency Channel Spacing
} 
& \colhead{HF Band Frequency Range}  
& \colhead{LF Band Frequency Range}  \\
 \colhead{(Frond-end Sideband)} 
 & \colhead{(MHz)}  
 & \colhead{(GHz)}  
 & \colhead{(GHz)}  
}
\decimalcolnumbers
\startdata
\multicolumn{4}{c}{B-08} \\
\hline
BB-1 (LSB)  & 0.9766 & 397.071 $-$ 398.946  & 127.051 $-$ 128.926  \\
BB-2 (LSB)  & 15.625 & 399.142 $-$ 400.876  & 129.122 $-$ 130.856  \\
BB-3 (USB)  & 15.625 & 409.184 $-$ 410.918  & 139.164 $-$ 140.898  \\
BB-4 (USB)  & 0.9766 & 411.114 $-$ 412.989  & 141.094 $-$ 142.969  \\
\hline
\multicolumn{4}{c}{B-09} \\
\hline
BB-1 (LSB) & 0.9766 & 657.041 $-$ 658.916   &  $\cdot\cdot\cdot$ \\
BB-1 (USB) & 0.9766 & 675.000 $-$ 676.875   &  $\cdot\cdot\cdot$ \\
BB-2 (LSB) & 0.9766 & 658.916 $-$ 660.791   &  $\cdot\cdot\cdot$ \\
BB-2 (USB) & 0.9766 & 673.125 $-$ 675.000   &  $\cdot\cdot\cdot$ \\
BB-3 (LSB) & 0.9766 & 655.124 $-$ 656.999   &  $\cdot\cdot\cdot$ \\
BB-3 (USB) & 0.9766 & 676.917 $-$ 678.792   & 152.167 $-$ 154.042 \\
BB-4 (LSB) & 0.9766 & 660.854 $-$ 662.729   &  $\cdot\cdot\cdot$ \\
BB-4 (USB) & 0.9766 & 671.188 $-$ 673.063   & 154.125 $-$ 156.000 \\
\hline
\multicolumn{4}{c}{B-10} \\
\hline
BB-1 (LSB) & 0.9766 & 884.629 $-$ 886.504 & $\cdot\cdot\cdot$ \\
BB-1 (USB) & 0.9766 & 905.954 $-$ 907.829 & $\cdot\cdot\cdot$ \\
BB-2 (LSB) & 0.9766 & 886.504 $-$ 888.379 & $\cdot\cdot\cdot$ \\
BB-2 (USB) & 0.9766 & 904.080 $-$ 905.955 & $\cdot\cdot\cdot$ \\
BB-3 (LSB) & 0.9766 & 888.379 $-$ 890.254 & $\cdot\cdot\cdot$ \\
BB-3 (USB) & 0.9766 & 902.205 $-$ 904.090 & 302.827 $-$ 304.702 \\
BB-4 (LSB)$^{\mathrm{a}}$ & 0.9766 & 890.254 $-$ 892.129 & $\cdot\cdot\cdot$ \\
BB-4 (USB) & 0.9766 & 900.330 $-$ 902.205 & 304.785 $-$ 306.660 \\
\enddata
\tablecomments{
$^{\mathrm{a}}$ 
The frequency range between 890.664 and 890.860 in the Band~10 BB-4 (LSB) was excluded from the continuum imaging because of the HCN maser line.  
}
\end{deluxetable*}

\clearpage
\newpage

\begin{longrotatetable}
\begin{deluxetable*}{ccccccc}
\tablenum{4}
\tablecaption{HF-LBC-2021 Go/NoGo check Before Starting the Experiment 
\label{tbl:04}}
\tablewidth{0pt}
\tablehead{
\colhead{Exp Code}
 & \colhead{Source}
 & \colhead{Band}
 & \colhead{PWV}
 & \colhead{Phase rms at $\nu_{\mathrm{HF}}$} 
 & \colhead{Excess Path Length rms}
 & Execution Block \\
 & 
 & \colhead{(Frequency (GHz))}
 & \colhead{(mm)}  
 & \colhead{(top quartile)} 
 & \colhead{($\mu$m) (top quartile)} 
 & \colhead{(uid://A002/)}  
}
\decimalcolnumbers
\startdata
B-08 
 & J0423-0120
 & Band~3 (103)
 & 1.01
 & $28^{\circ}$
 & 57
 & Xf0e992/X4d02  \\   
B-09
 & J0423-0120
 & Band~4 (140)
 & 0.29
 & $50^{\circ}$
 & 63 
 & Xf07268/X317e \\    
B-10 
 & J0423-0120
 & Band~7 (350)
 & 0.28
 & $63^{\circ}$
 & 59
 & Xf06573/X2087 \\    
\enddata
\end{deluxetable*}
\end{longrotatetable}


\begin{deluxetable*}{cccccc}
\tablenum{5}
\tablecaption{Flux Density and Spectral Index of J0522-3627 Measured 
in the ALMA Grid Source Monitor
\label{tbl:05}}
\tablewidth{0pt}
\tablehead{
\colhead{Exp Code} 
& \colhead{Frequency} 
& \colhead{Spectral Index} 
& \colhead{Calculated Flux Density} 
& \colhead{Measurement Epoch} 
& \colhead{Measurement Epoch} \\ 
\colhead{} 
& \colhead{(GHz)} 
& \colhead{} 
& \colhead{(Jy)} 
& \colhead{(Band 3)} 
& \colhead{(Band 7)} 
}
\decimalcolnumbers
\startdata
B-08 & 405.0 & $-0.331$ & 4.481  & 2021 Sep 21 and 26 & 2021 Sep 11 and 25 \\
B-09 & 667.0 & $-0.253$ & 4.390  & 2021 Sep 9 and 11  & 2021 Sep 10 \\
B-10 & 896.2 & $-0.250$ & 4.137  & 2021 Sep 9         & 2021 Sep 3 and 10 \\
\enddata
\end{deluxetable*}

\clearpage
\newpage

\begin{deluxetable*}{ccccc}
\tablenum{6}
\tablecaption{Elliptical Disk Fitting Result of the ($u$,~$v$) Visibility 
Data of R~Lep 
\label{tbl:06}}
\tablewidth{0pt}
\tablehead{
\colhead{Exp Code} 
 & \colhead{Flux Density}
 & \colhead{Position Offset$^{\mathrm{a}}$} 
 & \colhead{Source Size}
 & \colhead{Source Position}
 \\
 & \colhead{(mJy)}
 & \colhead{($\Delta\alpha\cos{\delta}$, $\Delta \delta$) (mas)}
 & \colhead{Major Axis $\times$ Minor Axis (mas)}
 & \colhead{Angle (deg)}
}
\decimalcolnumbers
\startdata
B-08 
 & 37.3 $\pm$ 0.1
 & ($-0.0$, $0.8$)
 &  14.2 $\times$ 14.2 
 & $\cdot\cdot\cdot$
 \\
B-09
 & 106.9 $\pm$ 0.5
 & ($-0.1$, $0.5$)
 &  16.3 $\times$ 15.6 
 & 31
 \\
B-10 
 & 179.9 $\pm$ 1.7
 & ($0.1$, $1.2$)
 &  18.1 $\times$ 18.1 
 & $\cdot\cdot\cdot$
 \\
\enddata
\tablecomments{
$^{\mathrm{a}}$ 
The origin is at the phase tracking center of R~Lep.   
}
\end{deluxetable*}

\clearpage
\newpage

\begin{deluxetable*}{ccccccc}
\tablenum{7}
\tablecaption{Image Characteristics of R~Lep with Application of B2B Phase Referencing Alone
\label{tbl:07}}
\tablewidth{0pt}
\tablehead{
\colhead{Exp Code}
 & \colhead{Data Type}
 & \colhead{Peak Flux Density}
 & \colhead{Image rms Noise}
 & \colhead{Synthesized Beam} 
 & \colhead{Synthesized Beam}
 & \colhead{Peak Brightness}
\\
 & 
 & \colhead{(mJy~beam$^{-1}$)}
 & \colhead{(mJy~beam$^{-1}$)}
 & \colhead{Major Axis $\times$ Minor Axis}
 & \colhead{Position Angle}
 & \colhead{Temperature (K)} 
\\
 & & 
 &
 & (mas)
 & (deg)
 &
}
\decimalcolnumbers
\startdata
B-08  
 & Continuum
 & 25.1 
 & 0.1
 & $13.6 \times 12.1$ 
 & $-42$
 &  1089  \\
B-09 
 & Continuum
 & 26.4 
 & 0.3
 & $6.6 \times 6.1$ 
 & $-52$
 & 1750 \\
B-10
 & Continuum
 & 23.0 
 & 0.8
 & $4.9 \times 4.7$ 
 & $-33$ 
 & 1491 \\
 B-10
 & HCN maser$^{\mathrm{a}}$
 & $735.2\times 10^{3}$ 
 & $7 \times 10^{3}$
 & $5.6 \times 4.9$ 
 & $-19$ 
 & $40.73\times10^{6}$ \\
\enddata
\tablecomments{
$^{\mathrm{a}}$ 
The original data cube with the velocity width of 0.3~km~s$^{-1}$ ($v_{\mathrm{_{\mathrm{LSRK}}}}=9.4$~km~s$^{-1}$). 
}
\end{deluxetable*}

\clearpage
\newpage

\begin{longrotatetable}
\begin{deluxetable*}{cccccccc}
\tablenum{8}
\tablecaption{Image Characteristics of R~Lep with Application of B2B Phase Referencing$+$Continuum Self-cal
\label{tbl:08}}
\tablewidth{0pt}
\tablehead{
\colhead{Exp Code}
 & \colhead{Data Type}
 & \colhead{Peak Flux Density}
 & \colhead{Image rms Noise}
 & \colhead{Synthesized Beam} 
 & \colhead{Synthesized Beam}
 & \colhead{Peak Brightness}
 & \colhead{Image}\\
 & 
 & \colhead{(mJy~beam$^{-1}$)}
 & \colhead{(mJy~beam$^{-1}$)}
 & \colhead{Major Axis $\times$ Minor Axis}
 & \colhead{Position Angle}
 & \colhead{Temperature (K)} 
 & \colhead{Coherence}\\
 & & 
 &
 & (mas)
 & (deg)
 &
 & (\%)
}
\decimalcolnumbers
\startdata
\multicolumn{8}{c}{B2B phase referencing$+$continuum self-cal} \\
\hline
B-08 
 & Continuum
 & 27.6 
 & 0.1
 & 13.7 $\times$ 12.2 
 & $-42$
 &  1187
 &  $\cdot\cdot\cdot$ \\
B-09
 & Continuum
 & 63.9 
 & 0.3
 & $10.8 \times  9.3$ 
 & $13$
 & 1715 
 & $\cdot\cdot\cdot$ \\
B-10 
 & Continuum
 & 170.0 
 & 1.5 
 & $19.5 \times 16.1$ 
 & $-47$ 
 & 802 
 & $\cdot\cdot\cdot$ \\
\hline
\multicolumn{8}{c}{B2B phase referencing alone with flags caused by continuum self-cal} \\
\hline
B-08  
 & Continuum
 & 25.5 
 & 0.1
 & $13.9 \times 12.4$ 
 & $-41$
 &  1060 & 92 \\
B-09 
 & Continuum
 & 52.9 
 & 0.4
 & $10.8 \times 9.3$ 
 & $13$
 & 1399 & 83\\
B-10
 & Continuum
 & 130.7 
 & 1.7
 & $19.5 \times 16.1$ 
 & $-47$ 
 & 617 & 77\\
\enddata
\end{deluxetable*}
\end{longrotatetable}

\clearpage
\newpage

\begin{longrotatetable}
\begin{deluxetable*}{cccccccc}
\tablenum{9}
\tablecaption{Image Characteristics of R~Lep with Application of B2B Phase Referencing$+$HCN Maser Self-cal
\label{tbl:09}}
\tablewidth{0pt}
\tablehead{
\colhead{Exp Code}
 & \colhead{Data Type}
 & \colhead{Peak Flux Density}
 & \colhead{Image rms Noise}
 & \colhead{Synthesized Beam} 
 & \colhead{Synthesized Beam}
 & \colhead{Peak Brightness}
 & \colhead{Image}\\
 & 
 & \colhead{(mJy~beam$^{-1}$)}
 & \colhead{(mJy~beam$^{-1}$)}
 & \colhead{Major Axis $\times$ Minor Axis}
 & \colhead{Position Angle}
 & \colhead{Temperature (K)} 
 & \colhead{Coherence}\\
 & & 
 &
 & (mas)
 & (deg)
 &
 & (\%)
}
\decimalcolnumbers
\startdata 
\multicolumn{8}{c}{B2B phase referencing$+$HCN maser self-cal} \\
\hline
B-10 
 & HCN maser$^{\mathrm{a}}$
 & $1247 \times 10^{3}$ 
 & $3 \times 10^{3}$
 & $5.4 \times 4.9$ 
 & $-21$
 & $114.54 \times 10^{6}$ 
 & $\cdot\cdot\cdot$ \\
B-10  
 & Continuum
 & 36.0 
 & 0.9
 & $5.3 \times 4.8$ 
 & $-27$
 & 2065 
 & $\cdot\cdot\cdot$ \\
\hline
\multicolumn{8}{c}{B2B phase referencing alone with flags caused by HCN maser self-cal} \\
\hline
B-10
 & HCN maser$^{\mathrm{a}}$
 & $757.5\times 10^{3}$ 
 & $7 \times 10^{3}$
 & $5.6 \times 5.0$ 
 & $-26$ 
 & $41.84\times10^{6}$ 
 & 61\\
B-10
 & Continuum
 & 25.4 
 & 1.0
 & $5.3 \times 4.8$ 
 & $-27$ 
 & 1457 
 & 70 \\
\hline
\enddata
\tablecomments{
$^{\mathrm{a}}$ 
The original data cube with the velocity width of 0.3~km~s$^{-1}$ ($v_{\mathrm{_{\mathrm{LSRK}}}}=9.4$~km~s$^{-1}$). 
}
\end{deluxetable*}
\end{longrotatetable}

\end{document}